\documentclass[useAMS,usenatbib,usegraphicx]{mn2e}

\title{Microlensing in phase space II: Correlations analysis}
\author[Tuntsov \& Lewis]
{A.V. Tuntsov\thanks{E-mail: tyomich@physics.usyd.edu.au} 
\& G.F. Lewis \thanks{E-mail: gfl@physics.usyd.edu.au} \\ 
A29, School of Physics, University of Sydney, NSW 2006, Australia}

\begin{document}

\date{Accepted 2006 April 6. Received 2006 April 4; in original form 2006 January 16}

\pagerange{\pageref{firstpage}--\pageref{lastpage}} \pubyear{2006}

\maketitle

\label{firstpage}

\begin{abstract}
Applications of the phase space approach to the calculation of the microlensing autocorrelation function are presented. The continuous propagation equation for a random star field with a Gaussian velocity distribution is solved in the leading non-trivial approximation using the perturbation technique. It is shown that microlensing modulations can be important in the interpretation of optical and shorter-wavelength light curves of pulsars, power spectra of active galactic nuclei and coherence estimates for quasi-periodic oscillations of dwarf novae and low-mass X-ray binaries. Extra scatter in the brightness of type Ia supernovae due to gravitational microlensing is shown to be of order up to $0.2^m$ depending on the extent of the light curves.
\end{abstract}


\section{Introduction}

Statistical approaches to gravitational microlensing have a long history that extends back to at least the mid 1980s (e.g., \citealt{paczynski}, \citealt{kayser}, 
\citealt{dw}) when it was realized that in general, a large number of stars contribute to the deflection field, and as a result, it is nearly impossible 
to model individual microlensing light curves in detail unambiguously \citep{schneiderweiss, katz}. Since then, a number of methods have been developed to 
calculate the statistical properties of microlensing light curves. Apart from a few early papers, the methods employed
are numerical -- ray-tracing \citep{kayser, wws, schneiderweiss} and image-tracking techniques \citep{lewisrt, wittrt} -- because they alone are capable of calculating the magnification probability distribution function over a large range of magnification values.

However, it was soon realized that analytic methods could help solve the complementary problem of correlations in the magnification values at nearby 
positions in a microlensing light curve. Based in part on the work of \citet{dw}, \citet{schneider1} developed a method to calculate the deflection angle
probability distribution function and successfully applied it in a determination of the flux autocorrelation function for a source observed through a single lens
plane for a variety of the parameters \citep{schneider2}. However, this method did not attract much further interest because of 
substantial technical difficulties in its implementation. Neindorf (2003) took a further major step forward by developing a simple approximation to the
correlated deflection probability distribution that could be applied to many interesting astrophysical situations.

In a companion paper (Tuntsov \& Lewis 2006, hereafter Paper I), we presented a similar approach in which a phase space (or intensity) 
description of the deflection process is used before projecting intensities onto observable quantities. The method makes clear some of the important 
approximations used in studies of gravitational lensing and is easily generalized to include many lens 
planes along the line of sight, which is especially interesting in the case of a cosmologically distributed population of lenses. To avoid arbitrary 
`slicing' of the space into individual lens planes, we have also proposed a continuous approximation to the propagation of intensity moments arguing
that for average quantities there is no conflict between the thin lens approximation and a 3D distribution of lenses; we call this `nearly' 3D lensing.

In Paper I we have also solved the problem of the propagation of first-order moments. However, this is of little interest by itself as the results
follow trivially from the conservation laws and the required symmetries of the initial conditions. In the present paper we apply these results
to the non-trivial problem of the propagation of second-order moments of the flux; traditionally, the most interesting projections are the autocorrelation
function and its Fourier transform -- the power spectrum.

In section 2, we review the results of Paper I in relation to this subject and present a general model for the initial conditions. In 
section 3, the deflection potential is calculated in the first non-trivial approximation for a random star field with a Gaussian distribution of 
velocities. The formal solution of the propagation equation using the perturbation technique is obtained in section 4, which is followed by the
calculation of the flux power spectrum for a point-like source in the leading order of the deflector density. For clarity, most of the calculations of this
section are presented in the Appendices.

Section 5 discusses the application of these results. First we consider modifications to the power spectra of periodic and quasi-periodic sources. 
We show that known microlensing populations of the Galaxy can introduce $\sim 0.1^m$-deep modulations in the pulse profiles 
of a kpc-distant pulsar and significantly affect coherence estimates for quasi-periodic oscillations of X-ray binaries and dwarf novae in the mHz to Hz frequency range.
Then, corrections to the autocorrelation functions of aperiodic sources are considered; in particular, we find that in the $0.01 - 0.1$ 
range of redshifts used for the calibration of type Ia supernovae brightness, the extra scatter due to gravitational lensing can amount to 
$0.04^m - 0.2^m$ depending on the light curve sampling time-scales. We conclude with a short list of refinements to the method that can increase 
the quality of our predictions.

\section{Equation for the second order moments}
In paper I we have considered the transformation of the statistical moments of the intensity field as light propagates through a universe
filled with gravitational lenses. The second-order moments are defined as the average value of the product of intensities
at a pair of positions in the phase space $\mathcal{M}^2\ni\left({\bf r}_1, {\bf r}_2, \balpha_1, \balpha_2, t_1, t_2\right)$, defined by the 
spatial position of a light ray in the lensing plane ${\bf r}_i$, its direction $\balpha_i$ and the moment of observation $t_i$, $(i=1,2)$:
\begin{equation}
M_2\left({\bf r}_1, {\bf r}_2, \balpha_1, \balpha_2, t_1, t_2\right)\equiv{\bf E}\, I\left({\bf r}_1, \balpha_1, t_1\right) I\left({\bf r}_2, \balpha_2, t_2\right) .\label{M2def}
\end{equation}
The expectation is taken with respect to an ensemble of realization of the intensity field; further discussion of this ensemble is given below.

It was shown in Paper I, that the conservation of surface brightness in gravitational lensing -- that is, the conservation of the intensity along the ray --
results in the linearity of transformation on $M_2$ as the light rays propagate from one lens plane to another and get deflected at them. It is therefore
convenient to consider $M_2$ as a vector in some linear function space $\mathcal{V}\left(\mathcal{M}^2\right)$. In particular, $M_2$ can be represented in any other 
functional basis; certain properties of the transformations often suggest particularly convenient bases.

Under the thin lens approximation it is possible to introduce two operators, termed deflection $\hat\Phi$ and focusing $\hat{F}$, which describe how 
the intensity moments are modified when the light rays are broken on the lens planes ($\hat\Phi$) and when they propagate freely between the planes ($\hat{F}$).
Then, we derived a Fokker-Planck-type equation to describe the continuous limit of this treatment, and to answer the question of how the intensity moments
behave as a function of depth $D$ along the line of sight:
\begin{equation}
\frac{\partial M_2}{\partial D}=\left(\frac{\mathrm{d}\hat{\Phi}}{\mathrm{d}D}+\frac{\mathrm{d}\hat{F}}{\mathrm{d}D}\right)M_2. \label{fokkerplanck}
\end{equation}
Here $\mathrm{d}\hat\Phi/\mathrm{d}D$ and $\mathrm{d}\hat{F}/\mathrm{d}D$ are effectively the generators of $\hat\Phi$ and $\hat{F}$, describing
the changes in $M_2$ due to these operators applied over an infinitesimally small distance $\mathrm{d}D$.

In the case of the second-order moments, the focusing operator in the above equation remains the same:
\begin{eqnarray}
\lefteqn{\frac{\mathrm{d}\hat{F}}{\mathrm{d}D}=-\sum\limits_{i=1,2}\left(\balpha_i\cdot\partial_{{\bf r}_i}+\frac{\balpha^2_i}{2c}\partial_{t_i}\right)} \label{focusing2} %\\
% & &  = \sum\limits_{i=1,2}\left(\mathrm{i}\partial_{\btau_i}\cdot\partial_{{\bf r}_i}+\frac{1}{2c}\partial^2_{\btau_i}\partial_{t_i}\right) \nonumber
\end{eqnarray}

Unlike the first-order case, there is a difference in the symmetry properties of the deflection operator. 
The thin lens and small angles approximations still apply, and therefore the deflection probability depends only on the deflection angle 
$\bbeta_i=\balpha'_i-\balpha_i$, and not on $\balpha_i$; this leaves the associated deflection operator diagonal (proportional to a $\delta$-function) 
in $\btau_i$ coordinates -- Fourier conjugates to $\balpha_i$. In the focusing operator, we simply replace $\balpha_i$ with $-\mathrm{i}\partial_{\btau_i}$.

At the same time, deflection probabilities are no longer entirely independent of temporal or spatial coordinates as the 
deflection angles and potential time delays of the two rays are correlated. Dependence on $t$ means that the associated deflection 
operator is no longer diagonal in the temporal frequency basis. However, recalling that the potential time delays 
turned to be unimportant in the observationally accessible range of frequencies, we will ignore them in the subsequent analysis 
and in doing so avoid a difficult integral term in the equation for $M_2$. Then, the deflection probability operator is diagonal 
in ${\bf r}_i, \btau_i, t_i$ basis.

In the stationary and spatially uniform case, probability densities for the light rays to be deflected by $\bbeta_i$ 
depend on the difference between the spatial and temporal coordinates associated with the deflections:
\begin{eqnarray}
\lefteqn{\hat{\Phi}\left({\bf r}_1, \btau_1, t_1, {\bf r}_2, \btau_2, t_2 \rightarrow {\bf r}'_1, \btau'_1, t'_1, {\bf r}'_2, \btau'_2, t'_2\right)=} \label{diagonal} \\
& & \Phi\left({\bf r}_1, {\bf r}_2, t_1, t_2, \btau_1, \btau_2\right)\delta(\btau'_1-\btau_1)\delta(\btau'_2-\btau_2)\nonumber \\
& & \times\delta({\bf r}'_1-{\bf r}_1)\delta({\bf r}'_2-{\bf r}_2)\delta(t'_1-t_1)\delta(t'_2-t_2), \nonumber
\end{eqnarray}
so the equation for the second moment assumes the following form:
\begin{eqnarray}
\lefteqn{\partial_DM_2\left(D; {\bf r}_1, {\bf r}_2, t_1, t_2, \btau_1, \btau_2\right)=}\label{Ec} \\
& &\hspace{-3mm} \Bigl(\mathrm{i}\partial_{\btau_1}\cdot\partial_{{\bf r}_1}+\mathrm{i}\partial_{\btau_2}\cdot\partial_{{\bf r}_2} + \frac{1}{2c}\partial^2_{\btau_1}\partial_{t_1}+\frac{1}{2c}\partial^2_{\btau_2}\partial_{t_2} \nonumber\\
& & \hspace{-1mm} + \Bigl.p_2\left({\bf r}_1, {\bf r}_2, t_1,  t_2, \btau_1, \btau_2\right)\Bigr) M_2\left(D; {\bf r}_1, {\bf r}_2, t_1, t_2, \btau_1, \btau_2\right) \nonumber
\end{eqnarray}
where $p_2$ is the derivative of $\Phi$ with respect to $D$.

To complete the statement of the problem, one also needs to specify the initial conditions $M_2|_{D=0}=M^0_2$. In the present
study, we will restrict ourselves to the case considered in Paper I: an isotropic source located at $D=-D_s$. Then, for
the first moment one has:
\begin{equation}
M^0_1({\bf k}, \omega, \btau)=f({\bf k}, \omega)\frac{\mathrm{i}c}{\omega D_s}\exp\left[-\frac{\mathrm{i}c}{2\omega D_s}\left(\btau+{\bf k}D_s\right)^2\right], \label{M01}
\end{equation}
where $f({\bf k}, \omega)$ is the Fourier transform of the source profile.

However, there is an important difference between the first and second order moments. In Paper I the ensemble over which
one should average to obtain the moments was not fully defined because it served as a conceptual tool in the general
derivation. Two alternatives were provided for the definition -- one either looks at independent patches of the sky and 
averages quantities of interest along different lines of sight or observes a certain source long enough to consider different
patches of the light curve as independent elements of the ensemble.

In this paper, we are more concerned with practical applications and for this reason we will stick to the latter definition
as it is closer to the situation encountered in observations. In this case, an internally consistent definition of the 
averages can only be provided for sources that are statistically stationary. In the first order, this leads to the trivial
result -- in the absence of a special point in time, all modes with $\omega\not=0$ have zero amplitude
and stay so because of the homogeneity of equations, while the $\omega=0$ mode is preserved due to conservation of energy;
even for the alternative definition using different sky patches, it is hard to think of any reason for a certain time position
to stand out, as noticed in Paper I.

This can be expressed in a formal way. Indeed, shifting the origin of the time axis by a constant $T$ results in a phase 
factor $\mathrm{e}^{\mathrm{i}\omega T}$ in front of the Fourier transform of a function $F(t)$:
\begin{equation}
F_T(\omega)=\mathrm{e}^{\mathrm{i}\omega T}F(\omega);
\end{equation}
the same applies to all averages. Therefore, in a statistically stationary case, when ${\bf E}_T F={\bf E} F$
for all $T$ one has
\begin{equation}
{\bf E}_T F(\omega)=\mathrm{e}^{\mathrm{i}\omega T}{\bf E}\, F(\omega) ={\bf E}\, F(\omega), \hspace{5mm}\forall T.
\end{equation}
Hence, ${\bf E}\, F(\omega)\equiv 0$ for all $\omega\not=0$.

Similarly, for the second order moments one has
\begin{equation}
{\bf E}_T F_2(\omega_1, \omega_2)=\mathrm{e}^{\mathrm{i}(\omega_1+\omega_2)T}{\bf E}\, F_2(\omega_1, \omega_2),
\end{equation}
and the only non-trivial statistically stationary components are those with $\omega_2=-\omega_1=\omega/2$; we presume $\omega\ge 0$ from now on. 
When $F(t)$ itself is a real quantity -- as is the case for intensity and flux -- $F(-\omega)=F^*(\omega)$, and the function
\begin{equation}
P_F(\omega)={\bf E}\, F(-\omega/2)F(\omega/2)={\bf E}\,|F(\omega/2)|^2 \label{pspectrum}
\end{equation}
is the power spectrum of $F$ oscillations and represents a Fourier transform of the autocorrelation 
function of $F$.

If one now assumes that the source brightens and dims synchronously over its entire spatial profile so that it can be separated into a temporal
and spatial parts:
\begin{equation}
f({\bf k}, \omega)=g({\bf k})h(\omega) \label{sepf},
\end{equation}
the initial conditions for the second moment of intensity distribution of an isotropic stationary source located at $D=-D_s$
can be written as:
\begin{eqnarray}
\lefteqn{M^0_2({\bf k}_{1,2}, \Omega, \omega, \btau_{1, 2}) = \delta(\Omega)\frac{4c^2 P(\omega)}{\omega^2D_s^2}\,\overline{g({\bf k}_1)g({\bf k}_2)} }\label{M02} \\
& & \hspace{10mm} \times \exp\left[\frac{\mathrm{i}c}{\omega D_s}\left(\left(\btau_1+{\bf k}_1D_s\right)^2-\left(\btau_2+{\bf k}_2D_s\right)^2\right)\right]\nonumber
\end{eqnarray}
where $\Omega=(\omega_1+\omega_2)/2$, $\omega=\omega_2-\omega_1$. The average $\overline{g({\bf k}_1)g({\bf k}_2)}$ is used because correlations may be present in the source profile.
%For a point-like source, $g({\bf k})=1$. 
From now on we will drop the factor $\delta(\Omega)$ 
which arises from the infinite extent of time axis and thus can be considered as a constant through regularization.
%Modes with different $\Omega$ are not mixed by the propagation equation as we will see soon. 

\section{Deflection potential}
In calculating the deflection potential $p_2$ we will again assume that the distribution of the deflectors is statistically
uniform and stationary and ignore any correlations between individual deflectors; we will also
assume Poissonian distribution for the number of deflectors in a lensing plane. In this case, the Markov-Holtsmark-Chandrasekhar \citep{chandrasekhar, neindorf}
method gives the following result:
\begin{equation}
p_2\left(\btau_1, \btau_2, {\bf r}, t\right)=\nu\left|\left|{\mathcal{L}}\right|\right|\left(p(\btau_1, \btau_2, {\bf r}, t)-1\right) , \label{p2}
\end{equation}
where $\nu$ is the spatial density of the deflectors, ${\bf r}={\bf r}_2-{\bf r}_1$, $t=t_2-t_1$, $\left|\left|\mathcal{L}\right|\right|$ is the area of the lensing plane, 
and $p$ is the Fourier transform of the deflection probability density with 
respect to the two deflection angles, calculated for a single deflector in the plane.

The probability density itself can be easily obtained since, for a single deflector of mass $m$, 
the deflection angles uniquely determine its positions $\brho_{1,2}$ with respect to the rays: $\brho_i-{\bf r}_i=m\bbeta_i/\bbeta_i^2$.
Therefore, for a uniform plane,
\begin{equation}
p(\bbeta_1, \bbeta_2, {\bf r}, t)=\frac{m^4}{\bbeta^4_1\bbeta^4_2}\left|\left|{\mathcal{L}}\right|\right|^{-1}\phi\left(m\left(\frac{\bbeta_2}{\bbeta^2_2} - \frac{\bbeta_1}{\bbeta^2_1}\right) + {\bf r}, t\right) \label{pbeta}
\end{equation}
Here the fraction is the Jacobian of the $\brho_{1,2}\rightarrow\bbeta_{1,2}$ transform, while $\phi(\brho, t)$ is the probability
density for the position change $\brho$ over a time interval $t$ -- essentially, the velocity distribution of the deflectors. The simplest
choice will be to assume a Gaussian distribution of these velocities with average ${\bf v}$ and dispersion $\sigma$:
\begin{equation}
\phi(\brho, t)=\left(2\pi\sigma^2 t^2\right)^{-1}\exp\left[-\frac{\left(\brho-{\bf v}t\right)^2}{2\sigma^2 t^2}\right] \label{Gaussianphi}
\end{equation}

However, Fourier transforming~(\ref{pbeta}) even in this simplest case is a very hard task.
We can, however, expand it in Taylor series in $\btau_{1,2}$ and restrict ourselves to the most important contribution in this
expansion. This is a necessary exercise given that the problem for~(\ref{Ec}) is in ten dimensions -- numerical methods
will not be easy to implement here unless we have some analytic understanding of the solutions.

From the normalization of the probability
density, the zeroth-order term in Taylor expansion of $p(\btau_1, \btau_2, {\bf r}, t)$ is unity and it cancels out in~(\ref{p2}). The coefficients in front of $\btau_1$ and $\btau_2$ are the average deflections of 
the two rays which equal zero for a uniform lensing plane. Therefore, the first term of the expansion is of the second
order:
\begin{eqnarray}
\lefteqn{p(\btau_1, \btau_2, {\bf r}, t) - 1=-\frac{1}{2}\Bigl[\left(\langle(\btau_1\cdot\bbeta_1)^2\rangle+\langle(\btau_2\cdot\bbeta_2)^2\rangle\right)\Bigr.} \label{sndodrp}\\
& & \hspace{21mm} +2\Bigl.\langle(\btau_1\cdot\bbeta_1)(\btau_2\cdot\bbeta_2)\rangle({\bf r}, t)\Bigr]+\overline{o}\left(\btau^2_1, \btau^2_2\right) , \nonumber
\end{eqnarray}
where the angled brackets correspond to averaging over the distribution~(\ref{pbeta}).

For symmetry reasons, the averages on the first line do not depend on ${\bf r}$ and $t$, and therefore they cannot depend 
on the direction of $\btau_{1,2}$ either. Thus,
\begin{equation}
\left(\langle(\btau_1\cdot\bbeta_1)^2\rangle+\langle(\btau_2\cdot\bbeta_2)^2\rangle\right)=\frac{1}{2}\langle\bbeta^2\rangle\left(\tau^2_1+\tau^2_2\right).	\label{doublefirstorder}
\end{equation}
The other average in~(\ref{sndodrp}) is responsible for the correlations. It can be rewritten in a simpler form
\begin{equation}
2\langle(\btau_1\cdot\bbeta_1)(\btau_2\cdot\bbeta_2)\rangle({\bf r}, t)=\langle\bbeta^2\rangle\btau_1^T\hat{C}_0\left({\bf r}, t\right)\btau_2, \label{correlationterm}
\end{equation}
where $\hat{C}_0$ is the correlation matrix
\begin{equation}
C_0^{ij}\left({\bf r}, t\right)\equiv 2\langle\bbeta^2\rangle^{-1}\langle\beta^i_1\beta^j_2\rangle\left({\bf r}, t\right) \label{Cdef}
\end{equation}
The average in the angled brackets can be found using the joint distribution function of $\bbeta_{1,2}$:
\begin{equation}
\langle\beta_1^i\beta_2^j\rangle({\bf r}, t)=\int\mathrm{d}^2\bbeta_1\mathrm{d}^2\bbeta_2\,p(\bbeta_1,\bbeta_2, {\bf r}, t)\,\beta^i_1\beta^j_2. \label{bibj}
\end{equation}
Later we will need the the Fourier transform of the correlation matrix as a function of $\bf r$ and $t$:
\begin{equation}
\hat{\tilde{C}_0}\left({\bf k}, \omega\right)\equiv\int\mathrm{d}^2{\bf r}\mathrm{d}t\,\frac{\exp\left[\mathrm{i}\left({\bf k}\cdot{\bf r}+\omega t\right)\right]}{(2\pi)^3} \hat{C}_0\left({\bf r}, t\right) \label{C0ftdef}
\end{equation}
For a Gaussian distribution of velocities~(\ref{Gaussianphi}) this integral can be done 
analytically. One first integrates with respect to ${\bf r}$ (or equivalently, ${\bf r}-m{\bf \Delta} 1/\bbeta$) over an
infinite plane to obtain a Gaussian-type integral with respect to $t$, while the integrals with respect to $\bbeta_{1,2}$ 
decouple. As a result, one obtains a degenerate matrix:
\begin{eqnarray}
\lefteqn{||\mathcal{L}||\langle\bbeta^2\rangle\hat{\tilde{C}_0}({\bf k}, \omega)=\tilde{C}({\bf k}, \omega)\, {\bf e}_\alpha{\bf e}^T_\alpha} \label{Cft}
\end{eqnarray}
where $\alpha$ marks the direction of ${\bf k}$: ${\bf k}={k}{\bf e}_\alpha$ and
\begin{eqnarray}
\lefteqn{\tilde{C}({\bf k}, \omega)=\sqrt{\frac{2}\pi}\frac{m^2}{\sigma k^3}\exp\left[-\frac{\left({\bf v}\cdot{\bf k}+\omega\right)^2}{2\sigma^2 k^2}\right]    } \label{Cftdef} \\
& & \hspace{21mm}\times \left[J_0\left(\rho_0 k\right/2) -J_0\left(\Lambda \rho_0 k/2\right) \right]^2  ,\nonumber
\end{eqnarray}
where we have assumed that the upper limit $m/\rho_0$ on $\beta$ is determined by the lens physical size $\rho_0$ while the lower one,
$\rho_0/R$ comes from the extent of the lensing plane $R=\Lambda\rho_0$. For relativistic compact objects, such as black holes and neutron stars, $\rho_0/m\sim\mathcal{O}(1)$ while for white dwarfs $\rho_0/m\sim\mathcal{O}(10^3)$.
%where for the lower and upper limits of $\beta$ values of $m/R=1/\Lambda$ and $1$, respectively, have been assumed.

%For the above choice of $\phi$ it is also symmetric:
%\begin{equation}
%\hat{C}_0=\left(\matrix{c_\kappa+c_\gamma\cos 2\gamma & c_\gamma\sin 2\gamma \cr c_\gamma\sin 2\gamma & c_\kappa - c_\gamma\cos 2\gamma}\right). \label{kappagammadef}
%\end{equation}
In the case of continuous distribution of the deflecting surface density $\Sigma(\brho, t)$, the deflection angle power spectrum 
can be related directly to the statistical properties of $\Sigma$. For a spatially and temporally uniform case one has ($\Sigma$ is in mass per length squared units):
\begin{equation}
\langle\beta^i_1\beta^j_2\rangle ({\bf k}, \omega) = \frac{4 G^2}{c^4}\frac{k^ik^j}{k^4}\,P_\Sigma\left({\bf k}, \omega\right), \label{continousdensity}
\end{equation}
where $P_\Sigma$ is the projected density power spectrum, which can be found using an appropriate form of the Limber's equation. In the formula
above we have neglected retardation effects, therefore it is valid when the motion of deflecting matter is slow compared to the speed of light.

However, the smooth-matter description requires careful treatment of the correlations along the line of sight and will not be considered
in this paper.
%Whether the smooth-matter description can be incorporated into our continuous propagation approach is not clear, however -- 
%the obscure bit is how correlations between different lensing planes can be dealt with. We therefore proceed with the uncorrelated random star field.

\section{Formal solution and the flux power spectrum}

Neglecting the higher-order terms in~(\ref{sndodrp}), we may write~(\ref{Ec}) as
\begin{eqnarray}
\lefteqn{\partial_D M_2=-\frac{\nu}{2}||\mathcal{L}||\langle\bbeta^2\rangle\btau_1^T\hat{C}_0({\bf r}, t)\btau_2\,M_2 } \label{startperturbations}\\
& &\hspace{4mm} +\Bigl(\mathrm{i}\partial_{\btau_1}\cdot\partial_{{\bf r}_1}+\mathrm{i}\partial_{\btau_2}\cdot\partial_{{\bf r}_2} + \frac{1}{2c}\partial^2_{\btau_1}\partial_{t_1}+\frac{1}{2c}\partial^2_{\btau_2}\partial_{t_2}\Bigr)M_2 \nonumber \\
& &\hspace{4mm} -\frac{\nu}{4}||\mathcal{L}||\langle\bbeta^2\rangle\left(\tau^2_1+\tau^2_2\right)M_2\nonumber
\end{eqnarray}
In the absence of correlations represented by the first line of the right-hand side a dense system of eigenvectors $M_a$ (and corresponding eigenvalues $\lambda_a$) of the remaining operator
is provided by various products and associated sums of the form
\begin{eqnarray}
M_a({\bf r}_1, {\bf r}_2, \btau_1, \btau_2, t_1, t_2)=M^{(1)}_{a_1}({\bf r}_1, \btau_1, t_1)M^{(1)}_{a_2}({\bf r}_2, \btau_2, t_2)\label{unperturbed}
% & &\hspace{29mm} M^{(1)}_{a_1}\left({\bf r}_1, \btau_1, t_1\right)M^{(1)}_{a_2}\left({\bf r}_2, \btau_2, t_2\right),  \nonumber
\end{eqnarray}
\begin{equation}
\lambda_a=\lambda^{(1)}_{a_1}+\lambda^{(1)}_{a_2} \label{unperturbedlambda}, 
\end{equation}
with a composite index $a=(a_1, a_2)$, where $M^{(1)}_{a_i}$, $\lambda^{(1)}_{a_i}$ are the solutions for the first-order equation:
\begin{eqnarray}
\lefteqn{\lambda^{(1)}_{a_i}M^{(1)}_{a_i}\left({\bf r}, \btau, t\right)=\Bigl(\mathrm{i}\partial_\btau\cdot\partial_{{\bf r}}+\frac{1}{2c}\partial^2_\btau\partial_t\Bigr)M^{(1)}_{a_i}\left({\bf r}, \btau, t\right)} \label{firstordersolutions}\\
& &\hspace{20mm} -\frac{\nu}4||\mathcal{L}||\langle\bbeta^2\rangle\tau^2 M^{(1)}_{a_i}\left({\bf r}, \btau, t\right). \nonumber
\end{eqnarray}
As was found in Paper I, these represent harmonic modes in ${\bf r}$ and $t$ multiplied by an angular functions $\zeta'_{mn}(\btau)$ depending 
on the spatial and temporal frequencies ${\bf k}$, $\omega$ as parameters (see Appendix~\ref{basisappendix}). 

The functions $M^{(1)}_{a_1}$ form a complete orthonormal basis of the solutions, while their products of
form~(\ref{unperturbed}) do the same job for the problem considered in this paper. It is therefore convenient to rewrite~(\ref{startperturbations})
in the basis of $M_a$. Namely, let
\begin{eqnarray}
\lefteqn{M_2(D; {\bf r}_{1,2}, \btau_{1,2}, t_{1,2})=\sum\limits_a c_a\,\mathrm{e}^{\lambda_a D} M_a({\bf r}_{1,2}, \btau_{1,2}, t_{1,2}) } \label{stpinMa}
%& &\hspace{16mm}  \nonumber
\end{eqnarray}
with $a=({\bf k}_{1,2}, \omega_{1,2}, m_{1,2}, n_{1,2})$ running through all possible values.
When there are no correlations a set of $c_a=\mathrm{const}$ chosen such as to represent the initial conditions at $D=0$ gives a solution of~(\ref{startperturbations}). 

In the presence of correlations, which we will treat as a perturbation, we allow these `constant coefficients' 
to vary with $D$ to obtain the following system of linear equations in the perturbed case:
\begin{equation}
\dot{c}_{a}=\sum\limits_{a'} C_{aa'}\mathrm{e}^{(\lambda_{a'}-\lambda_a)D}\,c_{a'}   \label{coefseq}
\end{equation}
where the dot denotes differentiation with respect to $D$. The matrix elements $C_{aa'}$ are defined as
\begin{equation}
C_{aa'}\equiv-\frac{\nu}{2}||\mathcal{L}||\langle\bbeta^2\rangle \left(M^\dag_{a}, \btau_1^T\hat{C}_0\btau_2 M_{a'}\right). \label{matrixelements}
\end{equation}

%where the scalar product is given by a ten-dimensional integral with respect to all six variables 
%${\bf r}_{1,2}$, $\btau_{1,2}$, $t_{1,2}$. They are calculated in the appendix; let us, however, observe
%one important property relating to the structure of this matrix here. One can equally rewrite $M_a$ as functions of central 
%and relative coordinates ${\bf r}_{1,2}={\bf R}\mp{\bf r}/2$, $t_{1,2}=T\mp t/2$ and replace the subindices ${\bf k}_{1,2}$, 
%$\omega_{1,2}$ with ${\bf K}$, $\bf k$, $\Omega$, $\omega$. Then, given that, due to the symmetry, $\hat{C}_0$ is independent 
%of $\bf R$ and $T$, $C_{aa'}$ are zero at ${\bf K}\not={\bf K}'$ or $\Omega\not=\Omega'$ and therefore different harmonic modes
%in $\bf R$, $T$ propagate independently of each other.

The perturbation method then proceeds by introducing a dummy parameter $\epsilon$ to characterize the order 
of perturbation $\epsilon C$ and expanding $c_a(D)$ in its power series:
\begin{equation}
c_a(D)=c^0_a(D)+\epsilon c^1_a(D)+\epsilon^2 c^2_a(D)+ ... \label{serieseps}
\end{equation}
such that $c^0_a(D)$ functions correspond to no correlations at all, $c^1_a(D)$ are proportional to the first 
power of perturbation and so on. A sequence of successive corrections to $c_a(D)$ can now be obtained by equating 
coefficients at common powers of $\epsilon$:
\begin{equation}
\dot{c}^0_{a}=0, \label{pertzeroth}
\end{equation}
\begin{equation}
\dot{c}^1_{a}=\sum\limits_{a'}  C_{aa'} \mathrm{e}^{(\lambda_{a'}-\lambda_{a})D}\,c^0_{a'} \label{pertfirst}
\end{equation}
%\begin{equation}
%\dot{c}^2_{a'}=\sum\limits_a c^1_a e^{(\lambda_a-\lambda_{a'})D} C_{a'a} \label{pertsecond}
%\end{equation}
and so on.%; after solving them one either sums the resulting series or restricts oneself to its first few terms.

From the first of the above equations $c^0_{a}=\mathrm{const}$ and if we choose these constants to represent 
the initial conditions $M^0_2$ at $D=0$: 
$c^0_{a}=\left(M^\dag_a, M^0_2\right)$, the initial conditions in all other orders will be uniform 
\begin{equation}
c^i_a(0)=0, \hspace{5mm} i\in \mathcal{N} .\label{pertinitial}
\end{equation}
Physically, the zeroth-order term describes just the partial loss of the phase coherence in different modes as a result of geometric time delays
due to uncorrelated random deflections, as explained in Paper I.

Further, we restrict ourselves to the first non-trivial term in the expansion~(\ref{serieseps}). It is proportional to the strength of correlations in
the deflection angle at different points and describes how these correlations translate across to the intensity field. By mixing the intensity
modes of one-ray statistics $c^{(1)}_{a_1}c^{(1)}_{a_2}\exp[(\lambda_{a_1}+\lambda_{a_2})D]$ in a correlated way, extra correlations in the combined
field are created, which can be described by solving~(\ref{pertfirst}) using $c^0_a=\mathrm{const}$:
\begin{equation}
\mathrm{e}^{\lambda_a D} c^1_a(D)=\sum\limits_{a'} c^0_{a'} \int\limits_0^D\mathrm{d}D'\, C_{aa'} \mathrm{e}^{\lambda_a(D-D')} \mathrm{e}^{\lambda_{a'}D'} . \label{c1a}
\end{equation}
%The first factor on the left-hand side reflects the damping of this signal due to the same partial loss of the phase coherence. 
The power in the first-order
term is also turned into extra signal by correlated deflections, but the higher-order contributions will be neglected in this study.

Although the intensity field provides a complete description in the sense discussed above, the intensity is rarely observed
in the real life due to the limited angular resolution of our telescopes. The quantity that is normally measured is the flux $F$. 

$F$ is a linear functional on the intensity field and, just like intensity, defined for every spatial and time coordinate ${\bf r}, t$. 
Thus, in calculating it, one only needs to consider the projection in the angular sector -- namely, consider $\btau=0$. We can therefore prescind from the 
{\it exact} parametrization of ${\bf r}$ and $t$ dependence -- e.g., the temporal Fourier harmonics of the flux observed at points ${\bf r}_{1,2}$
\begin{eqnarray}
\lefteqn{F_2({\bf r_1}, {\bf r}_2, \omega_1, \omega_2)\equiv}\label{F2def} \\
& & {\bf E}\int\mathrm{d}t_1\mathrm{d}t_2\,\frac{\exp[-\mathrm{i}(\omega_1 t_1+\omega_2 t_2)]}{2\pi}F({\bf r}_1, t_1) F({\bf r}_2, t_2) \nonumber
\end{eqnarray}
will be those of the intensity projected at $\btau=0$ without any additional transformation.

In the approximation we use, there are two contributions to these quantities, associated with the zeroth and first-order terms in the perturbation
theory expansion. They both depend linearly on the initial conditions and therefore one can introduce two $D$-dependent functionals 
$F_2^{0,1}\in\mathcal{V}^*(\mathcal{M}^2)$ that calculate the flux Fourier components as functions of $M_2^0$ at respective accuracy.

In the Appendix~\ref{fluxfunctionalsappendix}, these functionals are calculated with respect to $M_a$ functions. Since they form an eigenbasis of the unperturbed problem,
$F^0$ is diagonal in this representation. This is not the case for $F^1$: correlated deflections can produce the power in the flux Fourier
spectrum at frequencies where the flux power spectrum due to initial conditions is zero.

A stationary, isotropic and synchronous source is defined by just two functions: the average source spatial profile Fourier transform 
$\overline{g({\bf k}_1) g({\bf k}_2)}$ and fluctuations power spectrum $P(\omega)$. Therefore the expectation value of the autocorrelation 
function Fourier transform (${\bf k}_{1,2}={\bf K}\mp{\bf k}/2$):
\begin{equation}
F^{0,1}_2({\bf K}, {\bf k}, \omega; D)=\sum\limits_{d'} F^{0,1}_{dd'}(D) P(\omega')\overline{g({\bf k'}_1)g({\bf k'}_2)} \label{Fdddefbody}
\end{equation}
with $d=({\bf K}, {\bf k}, \omega)$; here we use the summation instead of the integration sign to avoid unnecessarily long notation. 

The coefficients $F^{0,1}_{dd'}(D)$ are given in the Appendix~\ref{fluxfunctionalsappendix} as an explicit expression for $F^0_{dd'}$ and as an integral of an explicit
expression for $F^1_{dd'}$. They describe the fraction of variability power that, 
in the presence of correlated deflections on path length of $D$, is transferred from the original $(\omega', {\bf K}', {\bf k}')$-mode 
to a mode at $(\omega, {\bf K}, {\bf k})$. 
As noted above, $F^{0}$ is diagonal in this basis.

However, a question that is more interesting from the observational point
of view concerns the value of the flux temporal variability power $\mathcal{F}$ observed at a certain point ${\bf R}$ (or autocorrelation between a pair 
of points ${\bf r}_{1,2}={\bf R}\mp{\bf r}/2$) in the plane of the observer. This can be obtained by summation of all amplitudes multiplied by their Fourier mode value
at the given point(s).
\begin{eqnarray}
\lefteqn{\mathcal{F}^{0,1}(\omega, D)=\int\mathrm{d}\omega'\,\mathrm{d}^2{\bf k}_1\,\mathrm{d}^2{\bf k}_2\,\mathrm{d}^2{\bf k}'_1\,\mathrm{d}^2{\bf k}'_2\,F^{0,1}_{dd'}(D)} \label{F1rribody}\\
& & \hspace{8mm}\times\overline{ g({\bf k}'_1)g({\bf k}'_2)} P(\omega') \frac{\exp\left[-\mathrm{i}\left({\bf k}_1\cdot{\bf r}_1\right)-\mathrm{i}\left({\bf k}_2\cdot{\bf r}_2\right)\right]}{4\pi^2} \nonumber
\end{eqnarray}
For the non-trivial term, coefficients $F^1_{dd'}$ depend not only on $\omega, \omega'$ but spatial wave numbers as well, and therefore the summation
can only be done for a certain source profile $\overline{g({\bf k}'_1) g({\bf k}'_2)}$. 

For the purpose of illustrating this method we choose the simplest case of a point-like source at the origin 
of the source plane and a single observer at the origin of the plane of the observer ${\bf r}_{1,2}=0$.
To facilitate comparison with the results of Paper I, we assume that the source is located at $D=-D_s\le 0$ and that $D>0$ space is filled
with deflectors of constant number density $\nu$. We perform the calculations in Appendix~~\ref{psiappendix}, here only the necessary results are present.

In order to visualize the results we consider a `line profile broadening' function $\psi(\omega, \omega', D)$ that represents broadening of a single line in the power 
spectrum at a certain $\omega'$: $P(\omega)=P(\omega')\delta(\omega-\omega')$. $\psi$ is defined as the ratio between~(\ref{F1rribody}) and the value 
of the flux power spectrum $\left.\mathcal{F}^0(\omega', D)\right|_{\nu=0}$ 
which would have been observed at $D$ were the lensing population absent. The latter can be obtained by noticing that the power spectrum 
is quadratic in flux which obeys the inverse-squares law when the light rays are not deflected. For the above choice of the source profile and the position of the observer
(see the Appendix~\ref{psiappendix})
\begin{eqnarray}
\lefteqn{\mathcal{F}^0(\omega', D)=\frac{P(\omega')}{(D_s+D)^4}+\mathcal{O}\left(\nu^2\right) } \label{F0ddkappaobody}%\\
%& & \hspace{-7mm} \frac{4P(\omega')}{D_s^2\left(D_s+D\right)^2}\frac{\exp\left[-\frac\omega c\sqrt{\kappa(\omega)}D\right]}{\left|\alpha +1  - (\alpha-1)\exp\left[-\frac{\omega}c\sqrt{\kappa(\omega)}(1+\mathrm{i})D\right]\right|^2} \nonumber \\
%& & \hspace{-7mm} =\nonumber
\end{eqnarray}
and therefore

\begin{eqnarray}
\lefteqn{\psi^{0,1}(\omega, \omega', D)=(D_s+D)^4\int\mathrm{d}^2{\bf K}\mathrm{d}^2{\bf K}'\mathrm{d}^2{\bf k}\mathrm{d}^2{\bf k}'\,F^{0,1}_{dd'}(D)} \label{F1rrbody}
\end{eqnarray}

The part of $\mathcal{F}^0$ due to uncorrelated deflections is of the second order in the deflectors' density $\nu$. As shown in Paper I,
in all relevant astrophysical situations the density of deflectors is very low in terms of a dimensionless quantity $mD\sqrt{\nu\omega/c}$ that
controls the strength of gravitational lensing by a three-dimensional distribution of compact lenses.

The dominant non-trivial contribution to $\psi$ in this regime is supplied by the term due to correlated deflections 
$\mathcal{F}^{1}\sim\mathcal{O}(\sqrt{\nu})$. Assuming that parameters $\nu, m, \sigma, v$ are constant along the line of sight
one can introduce the frequency difference scale
\begin{equation}
\Delta\omega_0=\frac{2\sigma}{\rho_0} , \label{domega0defbody}
\end{equation}
an amplitude parameter
\begin{eqnarray}
\lefteqn{q=\sqrt{\frac{2\omega\kappa(\omega)}{3c}}\frac{2\left(D_s+D\right)^{3/2}}{\rho_0}} \label{qdefbody}\\
& &\hspace{-5mm} =\frac{m}{\rho_0}\sqrt{\frac{8\pi\ln\Lambda\,\nu\left(D_s+D\right)^3}{3}} \nonumber
\end{eqnarray}
and write down the following approximation for $\psi$ (we drop superscript `1' because this is a dominant non-trivial contribution):
\begin{eqnarray}
\lefteqn{\psi(\Delta\omega, D)\approx\frac{12\exp\left[-v^2/2\sigma^2\right]}{\sqrt{2\pi}\ln\Lambda\,\Delta\omega_0}\,\Psi\left(\frac{\Delta\omega}{\Delta\omega_0}, \frac{D_s}{D_s+D}, \frac{v}\sigma, q\right) }  \label{P1PPhibody} %\\
% & &\hspace{2mm} \frac{12\exp\left[-v^2/2\sigma^2\right]}{\sqrt{2\pi}\ln\Lambda\Delta\omega_0}\Psi\left(\frac{\Delta\omega}{\Delta\omega_0}, \frac{D_s}{D_s+D}, \frac{v}\sigma, \frac{m\Delta\omega_0}{2\sigma}\right) ,\nonumber 
\end{eqnarray}
where a scaled version of the line profile function
\begin{eqnarray}
\lefteqn{\Psi\left(y, s, \frac v\sigma, q\right)\equiv} \label{Psidefbody} \\
& & \hspace{-5mm} q^2\int\limits_0^\infty\mathrm{d}x\, w\left(s, xq\right)\exp\left[-\frac{y^2}{2x^2}\right] I_0\left(\frac v\sigma \frac y x\right)\left[J_0\left(x\right)-J_0\left(\Lambda x\right)\right]^2 .  \nonumber 
\end{eqnarray}
An auxillary function $w(s, \xi)$ is defined in the Appendix~\ref{psiappendix} and proportional to $\xi^{-3}$ for $\xi\gg 1$.
Since $\Lambda\gg 1$ we can safely neglect the second Bessel function in the square brackets when evaluating~(\ref{Psidefbody}) except for some extremely low values of $y$ when it does induce a small correction; $J_0$ can also be neglected in subsequent formulae~(\ref{Ydef}) and~(\ref{Y1def}).

\begin{figure}
\hspace{0cm} 
\includegraphics[width=80mm, angle=0]{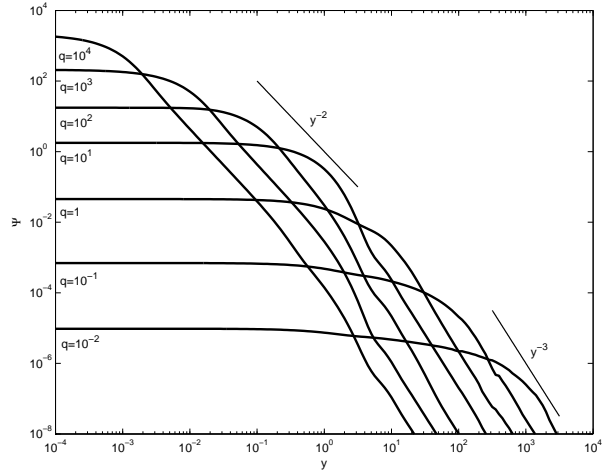}
\caption{The line profile function $\Psi(y, s, v/\sigma, q)$ as a function of frequency argument $y$ for $s=0$, $v/\sigma=1$ 
and a number of different values of the amplitude parameter $q$. The profile has a flat central part of the height $\propto q$ (for $q\ga\mathcal{O}(1)$) or $q^2$ (otherwise) extending to 
$y_c\sim\mathcal{O}(q^{-1})$ and wings falling off as $q^{-1} y^{-2}$ up to $y\sim\mathcal{O}(1)$ and as $q^{-1} y^{-3}$ afterwards. Certain `wiggles' can be seen in the shape of $\Psi$ which
stem from the behaviour of the Bessel function in~(\ref{Psidefbody}).}
\label{Psiofy}
\end{figure}

Figure~\ref{Psiofy} displays the behaviour of the line profile $\Psi$ as a function of the frequency difference variable $y$ for $s=0$ and $v/\sigma=1$. It has a flat central part and then falls off gradually. The behaviour of $\Psi$ is rather hard to study analytically as it has an essential singularity in the integrand; nevertheless, it can be understood qualitatively by noting that $\exp[-y^2/2x^2]$ effectively cuts out $x\la y$ from the integration region and using known form of $w(\xi)$ and Bessel functions.

There are two different regimes for $\Psi(y)$. At $s=0$, for $q\la \mathcal{O}(1)$ the height of the flat part is proportional to $q^{2}$ and it extends to $y\sim q^{-1}$ followed by a $\Psi(y)\sim q^{-1} y^{-3}$ tail afterwards:
\begin{equation}
\Psi(y)\sim\left[\matrix{ q^{2}, && y\la\mathcal{O}(q^{-1}) \cr  q^{-1} y^{-3}, && y\ga\mathcal{O}(q^{-1})}\right.,\hspace{7mm} q\la\mathcal{O}(1) . \label{Psilargezasymptotics}
\end{equation}
For $q\ga\mathcal{O}(1)$ the height of the central part is $\propto q$ and the extent is again $\mathcal{O}(q^{-1})$; however, the fall-off rate here is $\Psi(y)\sim q^{-1} y^{-2}$ up to $y\sim\mathcal{O}(1)$ where it turns to the same $q^{-1} y^{-3}$ tail: 
\begin{equation}
\Psi(y)\sim\left[\matrix{ q, && y\la\mathcal{O}(q^{-1}) \cr q^{-1} y^{-2}, && \mathcal{O}(q^{-1}) \la y\la\mathcal{O}(1) \cr q^{-1} y^{-3} && y\ga\mathcal{O}(1)} \right.,\hspace{3mm} q\la\mathcal{O}(1) . \label{Psismallzasymptotics}
\end{equation}
Oscillations of the Bessel functions enter the region not dominated by the exponential fall-off at different values of $y$ which results in a few `wiggles' in the resulting $\Psi(y)$ curve seen in Figure~\ref{Psiofy}.

In the opposite case of a distant source viewed through a localized overdensity of lenses, $1-s\ll 1$, the behaviour is generally the same but $q$ parameter should be replaced with its scaled version
\begin{equation}
q'=q(1-s)^{3/2}=\frac{m}{\rho_0}\sqrt{\frac{8\pi\ln\Lambda\,\nu D^3}{3}} ; \label{qpdef}
\end{equation}
the scaling arises from the fact that the main contribution to $w(s, \xi)$ comes, for $\xi\gg 1$, 
from the region of $t$ such that $(1-t)^2(t^2-s^3)\la\xi^{-2}$.

\begin{figure}
\hspace{0cm}
\includegraphics[width=80mm, angle=0]{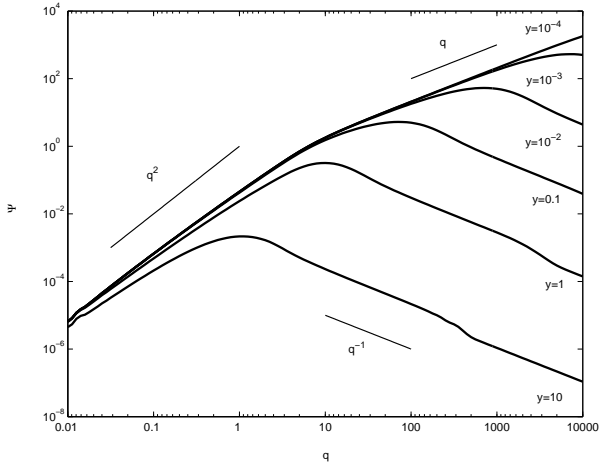}
\caption{The line profile function $\Psi(y, s, v/\sigma, q)$ as a function of the amplitude parameter $q$ for $s=0$, $v/\sigma=1$
and a number of different values of $y$. For every frequency argument value $y$, function $\Psi(q)$ reaches a maximum at
$q_c(y)$: $y_c(q_c)=y$ and then falls off as $y$ enters the wing of the line profile. The envelope curve of this
family represents the value of $\Psi$ near the edge of the central part $y_c(q)$.}
\label{Psiofz}
\end{figure}

Different cuts through $\Psi(y, q)$ can be obtained by considering $\Psi$ as a function of $q$, as shown in Figure~\ref{Psiofz}; here, again, we assume that the source is inside the deflectors-filled volume: $s=0$ and choose $v/\sigma=1$.
For a given frequency difference parameter $y$, $\Psi$ first increases ($y$ is a part of the central plateau) as either $\Psi\propto q^2$ for $q\la\mathcal{O}(1)$ or $\Psi\propto q$ for $q\ga\mathcal{O}(1)$, and then decreases as $q^{-1}$. The envelope curve of the $\Psi(q)$ family thus represents the height of the central part as a function of the amplitude parameter $q$. Taking into account the scaling~(\ref{qpdef}), $s\not=0$ (to be more precise, $1-s\ll 1$) curves can be obtained by a shift of the ones given in Figure~\ref{Psiofz} to the right.

For completeness, let us also state the dependence of the line profile on the velocity parameter $v/\sigma$. At $y\ll 1$ $\Psi$ is nearly
independent of $v/\sigma$ while at the shoulder and tail it scales as roughly $\exp[v^2/2\sigma^2]$. As a result, the shoulder and tail of the product $\exp[-v^2/2\sigma^2]\Psi(y, s, v/\sigma, q)$ are insensitive to the velocity while the centre of the profile
is significantly suppressed for $v/\sigma\gg 1$ and gently connected to the shoulder by a practically featureless curve.

The total fraction of the initial power added to the spectrum by correlated deflections can be obtained by integration of $\psi$ with respect to $\Delta\omega$:
\begin{eqnarray}
\lefteqn{\mathcal{Y}\left(D\right)\equiv\int\mathrm{d}\Delta\omega\,\psi\left(\Delta\omega, D\right)} \label{Ycaldef} \\
& &\hspace{2mm} =\frac{12\exp\left[-v^2/4\sigma^2\right]}{\ln\Lambda}I_0\left(\frac{v^2}{4\sigma^2}\right)Y\left(\frac{D_s}{D_s+D}, q\right), \nonumber
\end{eqnarray}
where
\begin{eqnarray}
\lefteqn{Y(s, q)\equiv\int\limits_0^\infty\mathrm{d}\xi\, \xi w\left(s, \xi\right)\left[J_0\left(\xi/q\right)-J_0\left(\Lambda \xi/q\right)\right]^2 } \label{Ydef} %\\
%& & \hspace{-7mm}\int\limits_s^1\frac{\mathrm{d}t\,t^2}{t^2-s^3}\exp\left[-\frac{z^2}{(t^2-s^3)(1-t)^2}\right]I_0\left(\frac{z^2}{(t^2-s^3)(1-t)^2}\right) . \nonumber
\end{eqnarray}

Another quantity of interest is the effective `line width' defined as:
\begin{eqnarray}
\lefteqn{\langle\Delta\omega\rangle\left(D\right)\equiv\mathcal{Y}^{-1}\int\mathrm{d}\Delta\omega\,\left|\Delta\omega\right|\,\psi\left(\Delta\omega, D\right)}  \label{domegamomentdef} \\
& & \hspace{7mm}=\Delta\omega_0\sqrt{\frac\pi 2}\,\frac{\exp\left[v^2/4\sigma^2\right]}{I_0\left(v^2/4\sigma^2\right)}\,\frac{Y_1(s, q)}{qY(s, q)} , \nonumber
\end{eqnarray}
where
\begin{eqnarray}
\lefteqn{Y_1(s, q)\equiv\int\limits_0^\infty\mathrm{d}\xi\, \xi^2 w\left(s, \xi\right)\left[J_0\left(\xi/q\right)-J_0\left(\Lambda \xi/q\right)\right]^2 ; } \label{Y1def} %\\
%& & \hspace{-7mm} z\int\limits_s^1\frac{\mathrm{d}t\, t^2}{\left(t^2-s^3\right)^{3/2}\left(1-t\right)}\, _2F_2\left(\frac 12, \frac 32; 1, 1; -\frac{2z^2}{\left(t^2-s^3\right)\left(1-t\right)^2}\right) \nonumber
\end{eqnarray}
we need to make use of the absolute value of $\Delta\omega$ in such a definition because as can be seen from the asymptotics
of $w(s, \xi)$ and Bessel function the second moment of $\Delta\omega$ does not exist.

Using~(\ref{Psilargezasymptotics}) one finds that, for $q\la\mathcal{O}(1)$, the total fraction of the power
created by correlated deflections $\mathcal{Y}$ is proportional to $q$ and is divided in roughly equal parts between the central plateau and $y^{-3}$ tail. They also make approximately equal contributions to the integral with respect to $y$ in~(\ref{domegamomentdef}), which is independent of $q$. As a result, the `effective line width' $\langle\Delta\omega\rangle\propto q^{-1}$.

For $q\ga\mathcal{O}(1)$, we find, using~(\ref{Psismallzasymptotics}), that $\mathcal{Y}$ saturates for $q\gg 1$; most of the power is split roughly equally between the central plateau that is high ($\propto q$) and narrow ($\propto q^{-1}$), and the $y^{-2}$ shoulder that extends out to $\Delta\omega\sim\mathcal{O}(\Delta\omega_0)$, only a small fraction $\sim\mathcal{O}(q^{-1})$ is in the $y^{-3}$ tail. The `effective line width' in this case $\langle\Delta\omega\rangle\propto\ln{q}/q$. 

\begin{figure}
\hspace{0cm}
\includegraphics[width=80mm, angle=0]{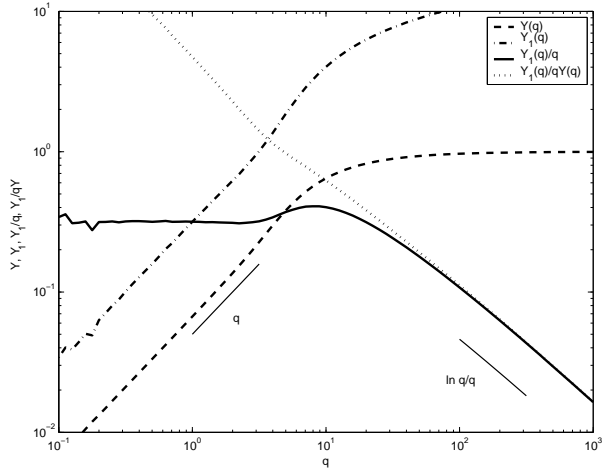}
\caption{Functions $Y(s, q)$, $Y_1(s, q)$, $Y_1(s, q)/q$ and the dimensionless effective `line width' $Y_1(s, q)/qY(s, q)$
as a function of the amplitude parameter $q$ for the source embedded in the deflectors-filled volume: $s=0$. $Y$ and $Y_1$ both saturate for large 
values $q\gg 1$ and have $\propto q$ tails for $q\ll 1$. The dimensionless effective `line width', on the contrary, is
a monotonically decreasing function of $q$.}
\label{YY1zY1}
\end{figure}

Figure~\ref{YY1zY1} shows the results of numerical integration in~(\ref{Ydef}, \ref{Y1def}). They agree well with the approximations above.

%As noted above, the wings and the central peak contribute approximately the same amount of power to modification of the power spectrum.

An important property of the broadening function $\psi$ in the low-density approximation is that it only depends on the frequency difference $\Delta\omega=\omega'-\omega$.
The convolution theorem then makes the transformation of the autocorrelation function (which, by virtue of the Wiener-Khinchine theorem is the Fourier
transform of the power spectrum) very simple: one just multiplies the autocorrelation function by the Fourier transform of $\psi$ and adds the result
to the original autocorrelation function. 
%However, before Fourier transforming $\psi$ we need to discuss observational limitations and ways in which we can model them. 

%The fact that $\Delta\omega_0$ is a negative power of $\nu$ and $D_s+D$ might seem strange suggesting that the width 
%of the line decreases where by common sense it is expected to increase. However, this simply reflects the fact that $\Psi$
%depends on $\nu$ in a more sophisticated way through its last argument $z$ and there is no simple scaling relation for
%this function. The effect, as measured by the value $\mathcal{Y}$, does increase when $z$ decreases. To quantify the 

%To simplify further analysis we place the source inside the deflectors-filled volume: $s=0$ and select a plausible value
%for systematic velocity of deflectors with respect to the observer $v/\sigma=1$. 

\section{Applications and discussion}

When discussing the application of the results we have obtained, it is important to interpret the results of observations correctly.
Within the model we have used, all real observational programs have two fundamental limitations. 

Firstly, it is the finite temporal resolution of observations $\tau$, making the information on variability at high frequencies inaccessible 
to observations. For the power spectra, this can be modelled by multiplying them by some window function of the frequency which decays rapidly for $\omega\ga\tau^{-1}$
and does not affect it for low frequencies. A good choice here could be a Gaussian normalized such that its value at zero is unity:
\begin{equation}
W_\tau(\omega)=\exp\left[-\frac{\tau^2\omega^2}2\right] .\label{Wtaudef}
\end{equation}
Another important limitation is the finite extent $T$ of any real observational program. This brings about an inability to estimate variations on very
long scales and results in a power spectrum where adjacent frequencies are no longer independent; as a result,
fine spectral features are smeared out. This can be modelled by convolving the power spectrum with a similar window function of width $\sim T^{-1}$; since the total power should
stay constant, the function must be normalized such that its integral is unity. An example of such a function is an integral-normalized Gaussian
\begin{equation}
W_T(\omega, \omega')=\frac{T}{\sqrt{2\pi}}\exp\left[-\frac{T^2(\omega'-\omega)^2}2\right] .	\label{WTdef}
\end{equation}

Let us now turn to some specific examples of the objects and study how their variability is affected by gravitational microlensing. We will first consider
periodic or nearly periodic sources, sketching the kind of effect gravitational microlensing brings about for objects observed through the microlensing population of
the Galaxy on distances of order 1~kpc; it will be convenient to use the frequency domain in this study. 

Then, aperiodic sources will be considered with a specific application to the extra scatter in the flux of standard candles; 
here the language of autocorrelation functions will be more appropriate. An important point regarding the standard candles is that they 
are most interesting on the cosmological scales while the analysis present in this work is essentially Newtonian in its use of space and time. In addition, the
parameters of the lensing population on these scales can only be guessed; therefore the results obtained in cosmological applications should be viewed 
cautiously but do give a general feeling of what the effect of microlensing on these scales should be.

In both cases optical (and shorter) wavelengths will be assumed, because at radio wavelengths interstellar scintillations are much more effective (e.g., \citealt{scalo, lazio}).

\subsection{Pulsars}

The most prominent features in the power spectra are displayed by objects which are periodic or close to being periodic. One of the most
interesting for us will be pulsars. The way their variability is affected is an interesting question on its own; in addition, they will
serve as a tool to help us interpret the action of spectral broadening on this simple example. 

The power spectra of periodic objects such as pulsars display an array of spectral lines at multiples of the spin frequency $\omega_p$; the
relative heights of these lines depend on the pulse shape. Broadening of every line is a typical feature of the multiplicative correction with
a slowly-varying function.

Indeed, let us consider a periodic source of flux
\begin{equation}
F_p(t)=\sum\limits_{n} A_n\mathrm{e}^{\mathrm{i}n\omega_p t}, \hspace{5mm} A_n=A_{-n}^* ,\label{fpdef}
\end{equation}
with a power spectrum
\begin{equation}
P_p(\omega)=\sum\limits_n \left|A_n\right|^2\delta(\omega-n\omega_p), \label{Ppdef}
\end{equation}
modulated by some function $1+\epsilon(t)$, with $\epsilon$ varying so slowly that its Fourier transform $\tilde\epsilon(\Delta\omega)$ is essentially 
contained in the region $\Delta\omega\ll\omega_p$.

Then, for the power spectrum of $F_p'(t)=F_p(t)(1+\epsilon(t))$ one has\footnote{We drop the infinitesimal correction to the discrete part of the spectrum.}
\begin{eqnarray}
\lefteqn{P_p'(\omega)=P_p(\omega)+\sum\limits_n\left|A_n\right|^2 P_\epsilon(\omega-n\omega_p)} \label{Pppdef} \\
& & \hspace{5mm}+\sum\limits_{n\not=m}A_nA_m^*\tilde\epsilon(\omega-n\omega_p)\tilde\epsilon^*(\omega-m\omega_p), \nonumber
\end{eqnarray}
where $P_\epsilon$ is the power spectrum of $\epsilon$: $P_\epsilon(\Delta\omega)=|\tilde\epsilon(\Delta\omega)|^2$.
The last line in~(\ref{Pppdef}) can be neglected due to assumption above, and comparing what is left to the results of our calculation in the previous section,
we can interpret $\psi(\Delta\omega)$ as the power spectrum of the magnification factor.

Let us estimate the quantities involved for a typical pulsar at a distance of $1\,\mathrm{kpc}$ observed through the population of white dwarfs of mass $0.5\,M_\odot$
with the space density of $0.01\,\mathrm{pc}^{-3}$ (as recent studies along different lines suggest -- \citealt{sion, hafizi, opennheimer, hangould}); assuming they have halo kinematics, 
we may choose $\sigma\approx 250\,\mathrm{km}/\mathrm{s}$ (e.g., \citealt{binneytremaine}). For such white dwarfs, $\rho_0/m\approx 3\times 10^{3}$ \citep{shapirot}. Given that pulsars generally have
high spatial velocities, a choice of $v/\sigma=2$ seems to be reasonable.

Then, for $\ln\Lambda=30$, the amplitude parameter becomes $q\approx 17$ and the frequency scale $\Delta\omega_0=0.05\,\mathrm{s}^{-1}$. Given a relatively
high value of the velocity parameter, $\psi$ starts off as an increasing function of $\Delta\omega$ as the latter changes from zero 
to $\Delta\omega_c\approx 0.01\,\mathrm{s}^{-1}$, where it turns to a shoulder followed by the $y^{-3}$ tail for $\Delta\omega\ga 0.3\,\mathrm{s}^{-1}$; a small
step is present at $\Delta\omega\approx 0.5\,\mathrm{s}^{-1}$.

The total extra power added to a line by gravitational microlensing is, according to~(\ref{Ycaldef}), $\mathcal{Y}\approx 15\%$ which is contributed mostly
by the shoulder -- the central part accounts for about a tenth of this power while the tail makes up a fraction of order few percent. The tail contributes 
slightly more to the integral~(\ref{Y1def}), and the effective line width, as given by~(\ref{domegamomentdef}), is $\langle\Delta\omega\rangle\approx 0.06\,\mathrm{s}^{-1}$.

These frequencies scale according to~(\ref{domega0def}) with parameter $\sigma$ -- e.g., the frequency difference scale would be twice as large if
the deflectors' velocity dispersion was only a half of the one assumed and twice as low if the deflector population was two times faster.

Our predictions for the effect of gravitational microlensing can be interpreted in a simple way for pulsars with spin frequencies greater than about $1\,\mathrm{s}^{-1}$.
Namely, we suggest that the pulse height is modulated by a slowly (and, presumably, randomly) varying function with an rms of $0.08$ stellar magnitudes
on the time-scales $T\ga \Delta\omega_c^{-1}\approx 100\,\mathrm{s}$; the time resolution required $\tau\la 1\,\mathrm{s}$. \label{sigmaestimate}

The actual pulsars show a degree of pulse-to-pulse variability that is much stronger than the expected $0.08^m$ contribution from lensing. However, unlike the lensing signal, these variations seem to be uncorrelated even on interpulse scale and therefore can be suppressed by averaging a large number of pulses. This does not necessarily mean a long time-scale -- the lensing signal correlation time is of order seconds. Therefore, {\it ceteris paribus}, the lensing signal is more likely to be detected in high-frequency pulsars.
 
%In accord with what has been said in the beginning of this subsection, consideration of a periodic emission is just a tool, though it simplifies the interpretation
%by allowing us to talk about the modulation of the signal amplitude.

\subsection{Quasi-periodic and related oscillations}

Another phenomenon of considerable interest in relation to the subject of this paper is quasi-periodic oscillations (QPOs) generally defined as broad -- up to
of order the central frequency -- lines seen in the power spectra of various astrophysical sources (e.g., \citealt{vanderklisbook}); more coherent features, with higher centroid frequency to
width ratios $Q$, are known by a variety of names, defined phenomenologically and often specific to a class of sources or even to the particular state of a given 
source. 

Within the Galaxy, these oscillations are associated with accretion onto compact objects -- black holes, neutron stars and white dwarfs -- from low-mass 
normal components. The frequencies of these oscillations span a wide range from mHz for cataclysmic variable stars that contain accreting white dwarfs 
to Hz and even kHz for low-mass X-ray binaries where the accretion proceeds onto neutron stars or black holes. Normally, features with $Q$ in the range of up to
a hundred or so are called QPOs, while for oscillations with greater coherence values terms such as normal, horizontal or flare branch oscillations (N/H/FBO) 
in the case of low-mass X-ray binaries \citep{vanderklisan, hasinger} and dwarf nova oscillations (DNOs), including longer-period ones (lpDNOs), 
in the case of cataclysmic variables are in use \citep{kluzniak, warnerlong, warner}. 
For convenience, we use the short term QPO more generally to mean all of the above when necessary.

It is believed that careful modelling of these spectral features can hold potential clues to the physical processes in the inner part of the accretion discs 
and that QPOs may become important probes of the strong gravitational and/or magnetic fields of the central compact objects \citep{psaltis, hellier, mauche}. Therefore, one has 
to interpret these features properly, including accounting for possible contamination of the signal en route to the observer; obviously, gravitational microlensing,
in the form described above, is one such possible `contaminant' able to significantly modify the intrinsic power spectrum -- although it does not change
the the centroid frequencies of QPOs, the total power in the features and their width can be considerably affected by lensing.

When making an estimate of the effect of lensing on the power spectra of these sources, the only change that seems necessary for the parameters of the 
lensing problem given in the previous subsection concerns the velocity -- a choice of $v/\sigma=1$ is probably more sensible in this case. As a result, the
central part is no longer suppressed but otherwise $\psi$ looks nearly the same. The total extra power added by lensing is slightly greater 
at $\mathcal{Y}\approx 0.25$, and the effective line width is lower: $\langle\Delta\omega\rangle\approx 0.035\,\mathrm{s}^{-1}$.

We can therefore conclude that gravitational microlensing has the potential to influence the total power observed in a given feature of the power spectrum 
and should be taken into account in their interpretation. With regards to coherence estimates, lensing makes a negligible contribution to the line width of kHz QPOs but cannot be ignored in
the interpretation of spectral lines at lower frequencies and may be the dominant effect in broadening lines at mHz frequencies. 

\begin{figure}
\hspace{0cm} 
\includegraphics[width=80mm, angle=0]{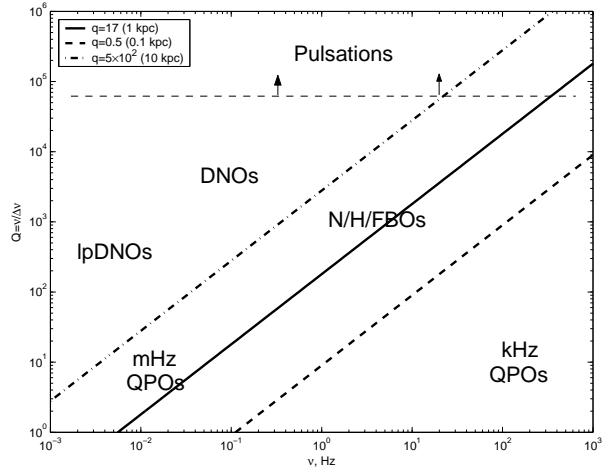}
\caption{A very rough classification of different spectral lines in terms of the central frequency $\nu$ and coherence parameter $Q=\nu/\Delta\nu$ 
($\Delta\nu$ is the spectral line width). The left part of the plot is occupied by different features seen in the power spectra of cataclysmic variables
while on the right one can see more rapid oscillations typical for accretion onto relativistic components. Gravitational lensing becomes important in 
the interpretation of coherence when broadening line width $\langle\Delta\omega\rangle$ becomes comparable to $\nu/Q$ -- these regions lie above the straight
lines seen in the figure which are plotted for $q=5\times 10^{2}$, $q=17$ and $q=0.5$ corresponding to the distances of 10 kpc, 1 kpc and 100 pc, 
respectively; the other parameters are fixed as in the text. The total extra power corresponding to these three values of $q$ are $0.32$, $0.25$ and $0.01$ --
that is, broadening contributes a little in the latter case although the line is formally quite broad.}
\label{fqplot}
\end{figure}

These results can be summarized in a graph that loosely classifies oscillations of different kinds in frequency-coherence coordinates (Figure~\ref{fqplot}). 
Apparently, whenever the lensing effective line width is of order $2\pi\nu/Q$ or greater and $\mathcal{Y}\sim\mathcal{O}(1)$, it should be taken into account. Therefore, lensing becomes important
for $Q\ga 2\pi\nu/\langle\Delta\omega\rangle$ -- the regions that lie above the straight lines in the plot.

\subsection{Aperiodic sources and standard candles}

Sources that do not show any prominent features in their power spectra, are more conveniently discussed in terms of the autocorrelation function. As mentioned
in the previous section, $\psi(\omega, \omega')$ only depends on the frequency difference $\Delta\omega=\omega'-\omega$ in the approximation we use. As a
result, the correction $\Delta f(t)$ to the autocorrelation function $f(t)$ can be found by multiplying $f(t)$ by the Fourier transform
of $\psi$:
\begin{equation}
\Delta f(t)=f(t)\tilde\psi(t) . \label{Deltaft}
\end{equation}

However, we should take into account the fact mentioned in the beginning of this section -- namely, limited temporal resolution of observations and thus the inability
to observe high frequency signal. To model this, we multiply $\psi(\Delta\omega)$ by the function $W_\tau(\omega)$ defined by~(\ref{Wtaudef}) and then
Fourier-transform the result to obtain an `observationally-corrected' version of $\tilde\psi$ (we include $\sqrt{2\pi}$ factor in the definition of $\tilde\psi$):
\begin{equation}
\tilde\psi_\tau(t)=\int\mathrm{d}\omega\,\mathrm{e}^{-\mathrm{i}\omega t} W_\tau(\omega)\psi(\omega) . \label{tildepsitaudef}
\end{equation}
Both $\psi$ and $W_\tau$ are even functions of $\omega$, so the exponent in the last expression can be replaced with the cosine function.	

Denoting $t_0\equiv\Delta\omega_0^{-1}$ we can rewrite the above formula in a form, similar to~(\ref{P1PPhibody}):
\begin{equation}
\tilde\psi_\tau(t)=\frac{12\exp[-v^2/2\sigma^2]}{\ln\Lambda}\tilde\Psi_{\tau}\left(\frac{t}{t_0},\frac{D_s}{D_s+D}, \frac{v}\sigma, q\right) \label{tildepsitautildePsi}
\end{equation}
with
\begin{eqnarray}
\lefteqn{\tilde\Psi_{\tau}\left(\eta, s, v, q \right)\equiv\int\frac{\mathrm{d}\omega}{\sqrt{2\pi}}\,\Psi\left(y, s, v, q\right) \mathrm{e}^{-\mathrm{i}y\eta-y^2\tau^2\Delta\omega_0^2/2}} \label{tildePsitaudef}
\end{eqnarray}

This integral also has to be done numerically. We have found the following rearrangement of the integral to produce fast and stable convergence for 
various values of the parameters (again, we assume $\Lambda$ is so large that the Bessel function it appears in can be neglected):
\begin{eqnarray}
\lefteqn{\tilde\Psi_{\tau}\left(\eta, s, v, q \right)= }\label{tildePsiintegral} \\
& & \hspace{14mm} \int\mathrm{d}x\,x\,w(s, x)\,J^2_0(x/q)\, j(\eta x/q, \tau\Delta\omega_0 x/q, v) \nonumber
\end{eqnarray}
with an auxiliary function
\begin{equation}
j(h, t, v)\equiv\sqrt{\frac 2\pi}\int\limits_0^\infty\mathrm{d}y\,\cos{hy}\,\exp\left[-\frac{1+t^2}2 y^2\right]\,I_0(vy). \label{jdef}
\end{equation}
At $v=0$, $q$ is simply a Gaussian of width $\sqrt{1+t^2}$ while at non-zero $v$, this function displays a few oscillations around zero value before being exponentially
suppressed on the same scale $\sim\mathcal{O}\left(\sqrt{1+t^2}\right)$.

Based on the similarity between $j(h)$ and a Gaussian, one can expect that $\tilde\Psi_\tau$ has an effective width of order $\langle\Delta\omega\rangle^{-1}$, which is approximately
$\sim t_0q/\ln{q}$ for $q\ga 1$. Similarly, we expect that $\tau$ should assume the role of this characteristic time when it exceeds the latter; 
cutting off some power at higher frequencies also leads to certain suppression in the value of $\tilde\Psi_\tau$ at low $t$. 

\begin{figure}
\hspace{0cm} 
\includegraphics[width=80mm, angle=0]{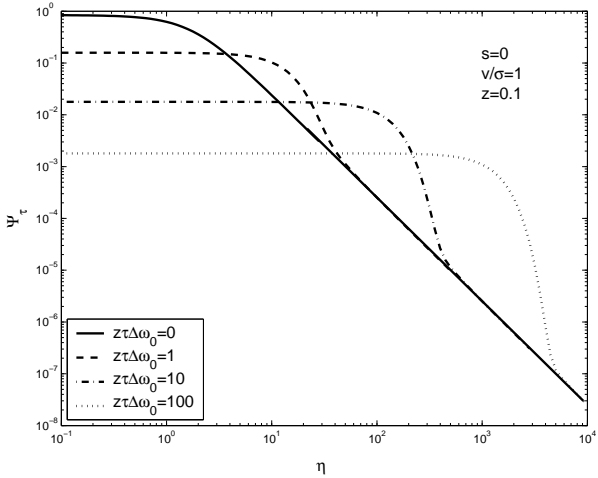}
\caption{The autocorrelation correction $\tilde\Psi_\tau$ as a function of $\eta=t/t_0$ for $q=10$, $s=0$, $v/\sigma=1$ and  a few values of the time 
resolution parameter $\tau/qt_0$ -- zero, one, ten and a hundred. It has a flat central part followed by $\eta^{-2}$ tail.}
\label{tildepsifig}
\end{figure}

\begin{figure}
\hspace{0cm} 
\includegraphics[width=80mm, angle=0]{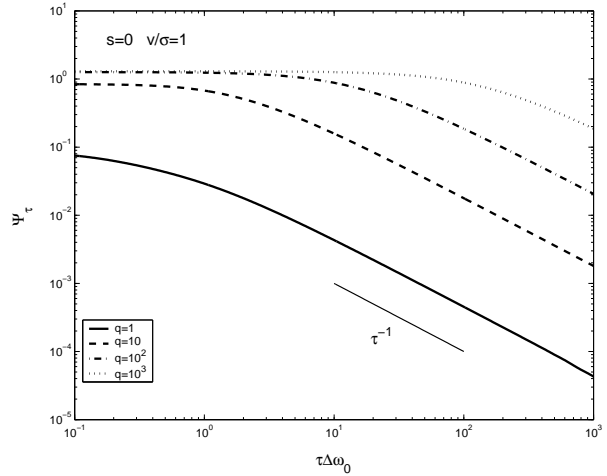}
\caption{The height of the central plateau $\tilde\Psi_\tau|_{\eta=0}$ as a function of the time resolution parameter $\tau/t_0$ for $q$ equal to one,
$0.1$, $0.01$ and $0.001$; the velocity parameter $v/\sigma=1$ and $s=0$. The quantity is weakly sensitive to the time resolution 
for $\tau\la\langle\Delta\omega\rangle^{-1}\approx t_0q/\ln{q}$ and decreases $\propto\tau^{-1}$ after that. 
%For the perfect resolution $\tau=0$ dependence on $z$ see Figure~\ref{YY1zY1} as $\tilde\Psi_0(0, s, v/\sigma, z)\equiv Y(0, s, v/\sigma, z).$
}
\label{tildepsiheightfig}
\end{figure}

\begin{figure}
\hspace{0cm} 
\includegraphics[width=80mm, angle=0]{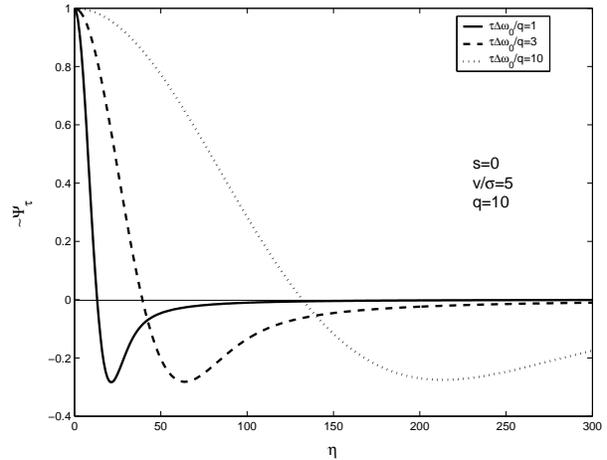}
\caption{Behaviour of the autocorrelation correction $\tilde\Psi_\tau$ as a function of $\eta$ for a high value of the velocity parameter $v/\sigma=5$ and
a number of different values of the time resolution parameter $\tau/qt_0$ -- one, three and ten; the amplitude parameter $q=10$ and $s=0$. To incorporate
both the different scales and the negatively-valued region, the linear scale is used but the functions are divided by their values at $\eta=0$. }
\label{tildepsihighvfig}
\end{figure}

Figure~\ref{tildepsifig} confirms these expectations. It shows the behaviour of $\tilde\Psi_\tau$ as a function of $\eta=t/t_0$ for $q=10$ and a few values 
of the 
parameter $\tau/t_0$ -- $0$, $10$, $10^2$ and $10^3$ that correspond to $\tau\langle\Delta\omega\rangle$ of zero, approximately one, ten and a hundred.
One can see a flat central plateau extending to approximately the greater of $\langle\Delta\omega\rangle^{-1}$ and $\tau$, followed by a $\eta^{-2}$ tail which
is nearly independent of $\tau$.
 The height of the plateau as a function of $\tau/t_0$ for a number of $q$ values is shown in Figure~\ref{tildepsiheightfig}.

In Figure~\ref{tildepsifig}, a small value of the velocity parameter $v/\sigma=1$ is assumed. For higher velocities the Fourier integral for $\tilde\Psi$ is dominated
by non-zero frequency modes, and $\tilde\Psi_\tau$ drops below zero outside the flat central part, then tends to zero value in the same, $\eta^{-2}$, manner 
from the negative side. Figure~\ref{tildepsihighvfig} illustrates this.

Although not easy, it is interesting to compare these findings to the results of previous works on calculation of the lensing autocorrelation function. Both 
Seitz~et~al. (1994) and \citet{neindorf} perform these calculations for a two-dimensional distribution of lenses, and the resulting curve depends on a number 
of parameters, including the effective convergence and shear and scaled radius of the source, while this work assumes a (nearly) three-dimensional 
distribution of lenses with no external shear, and calculations are performed for a point-like source. 

However, inspection of Figures 9 and 11 of Seitz~et~al. (1994) 
shows that their autocorrelation function $\Gamma(Y)$ does drop off as approximately an inverse square of $Y$ in the shearless and 
parallel-to-shear cases for $Y\ga 3$ while perpendicular-to-shear curve is clearly not a power law; it should be noted, however, that this impression is
based on measurements of the actually published plots. On the contrary, the growth of the autocorrelation function in Figure 1 of \citet{neindorf} seems to be
shallower than the inverse squares; however, it is the result of a best-fit procedure to the actual autocorrelation function observed in one of the quasars.

Let us now describe how an autocorrelation function of a distant source is affected by the transformation we have presented.
For any source, the intrinsic flux autocorrelation function $f(t)$ should tend to the square of the flux average value for large $t$. Let us for simplicity 
assume this value to be unity: $f(t)\rightarrow 1$ for $t\rightarrow\infty$. On top of that, one has all the information about second-order statistics of
variability, normally presenting itself as a decreasing -- not necessarily monotonically -- function of $t$. In particular, the $t=0$ value of the
autocorrelation function gives the flux dispersion: $\sigma^2=f(0)-1$.

When $f$ is subjected to the transformation described in this paper, the correction is proportional to $f(t)$, not the variable part $\bar{f}(t)=f(t)-1$, 
and as a result, even the constant brightness sources are affected:
\begin{equation}
\bar{f}'(t)=f'(t)-1=\tilde\psi(t)+\bar{f}(t)\left(1+\tilde\psi(t)\right) .\label{barfpt}
\end{equation}
As can be seen from~(\ref{tildepsitautildePsi}) and the results of the calculations in this section (e.g., Figure~\ref{tildepsiheightfig}), $\tilde\psi$
is less than unity for any realistic $\ln\Lambda$. Therefore, for mildly variable sources with $\bar{f}(0)=\sigma^2\la 1$, the above formula can be 
rewritten in the most simple form
\begin{equation}
\bar{f}'(t)\approx\bar{f}(t)+\tilde\psi(t) \label{barfptapprox}
\end{equation}
-- that is, the microlensing correction simply adds up to the intrinsic autocorrelation function.

The amount added is actually rather modest. For $\ln\Lambda=30$, the factor in front of $\exp[-v^2/2\sigma^2]\tilde\Psi$ in~(\ref{tildepsitautildePsi}) 
is 
approximately $0.41$ and $\tilde\Psi(0)$ saturates at about $1.3$ for $z\ll 1$. As a result, $\tilde\psi(t)\la 0.5$, and as with the intrinsic $\bar{f}(t)$,
the maximum is at $t=0$.

For sources with some memory of their past, one expects $\bar{f}(t)$ to be smooth at $t=0$ and therefore the $\tilde\psi(t)$ plateau is hard to spot. 
Whether gravitational lensing becomes a dominant contributor to $\bar{f}'(t)$ therefore depends largely on the
shape of the intrinsic function $\bar{f}(t)$. As $\tilde\psi\propto t^{-2}$ at large $t$, it necessarily becomes dominant if $\bar{f}(t)$ decreases as
a power law $t^{-n}$ with $n>2$ or more rapidly -- e.g., exponentially. The well-sampled observed autocorrelation functions are indeed often fit with 
exponentials but this is mostly for the purpose of simplicity and convenience (\citealt{hawkins96, devries04}). A simple transformation of the autocorrelation
function -- the structure function -- is more often used to characterize variability, but despite its simplicity, it is non-linear on $f(t)$ and therefore
is less suitable for our analysis.

The power spectra of various quasars and AGNs have also been studied observationally in a number of works \citep{giveonmaoz, paltani, fiore, edelson}. Traditionally, their 
variability is divided into a short and long time-scale; the former has stronger amplitudes and is thought to arise from the processes in the inner
engines of quasars while the origin of the long-term variability is less clear and a variety of different mechanisms have been proposed (e.g., \citealt{bregman, zackrisson});
it is quite likely, that a number of mechanisms actually contribute and microlensing could be just one of those. We therefore are only concerned with
the long-term variability.

According to the theory presented in this study, any strong signature in the power spectrum of a quasar will be broadened by a universal function $\psi(\Delta\omega)$.
In particular, if a quasar or AGN has only one dominating line at $\omega=0$, its entire power spectrum will be given by this function. 
According to estimates given in the previous section, for $q\gg 1$ most of the power resides in the $\Delta\omega^{-2}$ shoulder of $\psi$. The observed power spectra on long
time-scales are well described by a featureless power law with and index $\sim -2$ outside the central region where the spectrum flattens (it is hard to
estimate it here, though, due to the limited extent of the light curves). \citet{edelson} give an estimate of $-1.74 \pm 0.12$ for the slope of the
X-ray fluctuations power spectrum of a nearby Seyfert galaxy NGC 3516; the characteristic frequency of the break in the spectrum is $\omega_c\approx 2\pi/1\,\mathrm{month}\approx3\times 10^{-6}\,\mathrm{s}^{-1}$. \citet{giveonmaoz} report an average slope of $-2.0\pm 0.6$ (at frequencies range $10^{-7}\,\mathrm{s}^{-1}\la\omega\la 10^{-6}\,\mathrm{s}^{-1}$) for the sample of $\sim 40$ Palomar-Green quasars observed
regularly in the optical over seven years; based on the simulated sensitivity of their method, they conclude that they cannot reliably infer these
quantities for individual quasars in the sample. They do, however, report two cases with an apparent break in the power-law slope at a frequency $\omega_c\approx 3\times 10^{-7}\,\mathrm{s}^{-1}$.
However, estimation of the power law index is still difficult with the present-day data accuracy and coverage. The total variability power in these
spectra are of order a few to about twenty percent (rms).

Therefore, although it would be extremely preliminary to
 claim that these data support our predictions, they certainly do not rule out the microlensing hypothesis
at the present time. Some other aspects of this variability, such as chromaticity and (tentative) asymmetry of light curves, however, are more difficult to
accomodate within it \citep{zackrisson}. Nevertheless, it would be interesting to consider whether the microlensing-induced broadening, if ever detected,
can place meaningful constraints on the global microlensing population. The model presented in the previous section is probably too crude to describe the actual
distribution of microlensing bodies in a cosmological situation with strong clustering of lenses, and more work needs to be done to infer 
quantitative characteristics of this population from the available data.

Even if microlensing never shows up as a dominant component of the autocorrelation function or the power spectrum of a source, it still contributes significantly 
to the dispersion of the flux. Indeed, $\tilde\Psi(0)$ saturates very quickly with $q\ga 1$ (see Figures~\ref{YY1zY1} and~\ref{tildepsiheightfig}) and 
for the lensing population considered in this section, this happens relatively quickly, at distances of order a few kpc. As a result, for moderate values
of $v/\sigma$, one gets a top-up to the magnitude of order $0.1^m - 0.2^m$, as estimated in subsection 5.1.

In the language of frequencies, it is clear that this variability is spread over a range of timescales from roughly $t_0$ to $qt_0$ 
and consequently can be detected only if the time resolution $\tau$ of observations is better than $t_0$; a fraction of order $\tau/(qt_0)$ of this variability
will be `smoothed out' if $\tau\ga t_0$. The larger time-scale $T$ is responsible for the sample completeness -- for shorter light curves large variations
are likely to be left outside its coverage; this is not an issue if the dispersion of an ensemble of similar sources is considered, which is the case when we
calculate the extra scatter in the brightness of standard candles due to microlensing is.

Perhaps, the most interesting of such questions deals currently with the magnitude scatter of distant type Ia supernovae used in cosmology to measure
the luminosity distance as a function of redshift for constraining the evolution of the Universe \citep{riess98, perlmutter99}. The full lensing aspect of the problem is beyond
the scope of this paper (see, e.g., \citealt{holzlinder, ammanulah03, hui} for recent studies of the role lensing plays here) as the geometry of the Universe assumed here is 
flat and static and cannot describe the high-redshift Universe.

However, the treatment of this paper still applies to the Universe at moderate redshifts up to at least $0.01$ -- $0.1$, and this is precisely the redshift
range used to determine the zero-point of the redshift-magnitude relation and its uncertainty \citep{snf}; an estimate of the effect in cosmological settings should
also give some idea of what kind of time-scales are involved. 

The cosmological microlensing population is still quite
a mystery. For the economy of the hypotheses we will still assume that the deflectors are unseen white dwarfs of $0.5 M_\odot$ which are uniformly distributed
with density $\nu$. It can be written in terms of the fraction of the critical density in deflectors $\Omega_d$ and Hubble constant $H_0$ as:
\begin{equation}
\nu=\frac{3H_0^2\Omega_d}{2\pi c^2 m} , \label{nuOmega}
\end{equation}
which leads to
\begin{eqnarray}
\lefteqn{q=\frac{2D\sqrt{\Omega_d\ln\Lambda}}{D_H}\frac{\sqrt{mD}}{\rho_0} , } \label{zcosm}%\\
%& &\hspace{-4mm} \approx \frac{6.1\times 10^{-11}}{\sqrt{\Omega_d\ln\Lambda}} \left(\frac{M}{M_\odot}\right)^{1/2}\left(\frac{D}{1\,\mathrm{Gpc}}\right)^{-3/2} \nonumber
\end{eqnarray}
where the Hubble distance $D_H=c/H_0$ has been introduced; one can also recognize $\sqrt{mD}$ as the radius of the Einstein-Chwolson ring for a lens of mass $m$ seen from 
a distance $D$ \citep{schneiderbook}. $D/D_H$ is very close to the redshift $z$ for small values of the latter in
a flat Universe. For a currently preferred value $D_H\approx 1.4\times 10^{28}\,\mathrm{cm}$ \citep{lahavliddle}, one has
\begin{equation}
q\approx{4\times 10^{7}}\frac{\sqrt{\Omega_d\ln\Lambda}}{\rho_0/10^9\,\mathrm{km/s}}\left(\frac{M}{M_\odot}\right)^{1/2}\,z^{3/2} .\label{zcosmest}
\end{equation}

An estimate $\Omega_d\ln\Lambda\approx 1$ can be regarded conservative since we know that MACHOs contribute around ten to twenty per cent of the dark matter 
in our neighbourhood \citep{alcock2000, lasserre2000, sadoulet, evansbelokurov}, and the dark matter share in the composition of the Universe is $\Omega_{DM}\approx 0.27$ \citep{lahavliddle}. 
As a result, $q\gg 1$ at redshifts $z\sim 10^{-5}$ and above -- which includes all intergalactic scales. It also seems
sensible to choose the same velocity dispersion $\sigma=250\,\mathrm{km/s}$ (e.g., \citealt{tormen93, bahcall96}) and the velocity parameter $v/\sigma=1$.

For these parameters, $\left.\tilde\psi(0)\right|_{\tau=0}=\mathcal{Y}\approx 0.35$. The time-scales are $t_0=\Delta\omega_0^{-1}=20\,\mathrm{s}$ and
\begin{equation}
qt_0\approx 8\times 10^{8}\,\mathrm{s}\,\frac{\sqrt{\Omega_d\ln\Lambda}}{\rho_0/10^9\,\mathrm{cm}}\left(\frac{M}{M_\odot}\right)^{1/2}\,z^{3/2} . \label{largertimescale}
\end{equation}
At $z=0.01$ the latter time-scale is approximately $10$ days. The brightness of a supernova is normally derived from a fit to the observed light curve with the extent of order ten to fifty days \citep{riess04}, 
which, therefore, should be considered the time resolution. Given that $\tilde\psi_\tau(0)\propto\tau^{-1}$ at  these scales, 
we can expect the extra scatter in the magnitude to be
\begin{equation}
\sigma_m^{\mathrm{lens}}= \frac{\tilde\psi_\tau(0)}{2\cdot 0.4\ln 10}\approx 0.04^m - 0.2^m \label{scatterestimate}
\end{equation}
at these redshifts. Such an extra scatter is comparable to the intrinsic scatter in the brightness of type Ia supernovae, which is currently
estimated at $\sigma_m\sim 0.15^m$ \citep{tonry03, snls}, and therefore should
 be taken into account.

At greater $z\sim 0.05$, which is roughly the target redshift of the SuperNova Factory project \citep{snf}, the larger timescale becomes 
$t_0 q\approx 10^2\,\mathrm{days}$ and the whole extent of the high-quality light curve typically used in estimating its brightness 
$\tau\la 50\,\mathrm{days}$ is not sufficient to significally alter the scatter due to gravitational lensing. Therefore, this project should be able
to put some constraints on the spatial density of lenses $\nu$. In these settings, we cannot expect these constraints to be particularly useful because
most of $\sigma_m^{\mathrm{lens}}$ comes from the first $\sim 10 - 100\,\mathrm{kpc}$ of the optical path. If a method could be devised to actually
estimate the autocorrelation function from these data, it has the potential to serve such a probe.  However, as mentioned above, other more stable objects 
seen to high redshifts (e.g., quasars) could be better targets for such studies. The numerical predictions for the amplitude of the autocorrelation function correction
and the time-scales involved are, of course, the same as for the supernovae as long as both can be considered point-like.

\subsection{Conclusions}

In the present paper we have calculated the autocorrelation function of the flux of a point-like source through a population of uniformly distributed 
point-like gravitational microlenses using the lowest nontrivial order approximation for the deflection potential defined by~(\ref{p2}). In the last
section we have also identified several potential applications in illustration of the method. Although, given the simplicity of the approximation, 
the estimates we obtained should be taken rather cautiously, they do show that the effect is important and cannot be ignored.

It is sensible to conclude by recapping briefly the major assumptions used in this study, which should give us the idea of its applicability and show 
how and in which directions it can be improved. The `physical' approximations employed here are the use of the geometric optics and thin lens approximation 
as well as neglecting the potential time delays when writing down the propagation equation~(\ref{Ec}).
The first of these was shown at the very beginning to be an excellent one in most situations in the gravitational lensing (e.g., Schneider~et~al. 1992), 
the other two were extensively discussed in Paper I and Section~2 of the present paper.

The major mathematical approximations are the truncation of Taylor series in~(\ref{p2}) on the first non-trivial, correlation, term and the perturbation 
technique used in the solution of the propagation equation. The justification for series truncation comes from the fact that coefficients in front of the
higher-order powers are all proportional to the appropriate powers of the characteristic deflection angles, which only become of order unity very close to 
the lens; this never occurs for the deflectors considered in this paper; however, if one wants to consider black holes as deflectors, it will be necessary 
to look at these matters more carefully. The use of the perturbation technique can only be justified retrospectively by comparing corrections to the 
uncorrelated propagation of the quantities involved. Here, $\mathcal{F}^1$ indeed turned out to be not more than about $0.5$ of the zeroth-order quantity.
If, for a different problem, this is not the case, one would have to sum the entire series~(\ref{serieseps}) or seek other solution methods.

We have also only considered a point-like source and constant density of the deflectors. It does not pose any conceptual difficulties to perform the same 
calculations for a different source profile and deflectors density. However, such calculations are likely to require considerable numerical effort.

Apart from these approximations, we have also made certain assumptions about different parameters of the lensing population and most of our numerical 
predictions would change if very different parameters were considered. However, there are some generic features which do not suffer from these uncertainties,
the most robust of them are the shape of the power spectrum broadening function $\psi(\Delta\omega)$ and the autocorrelation function correction $\tilde\psi(t)$.
In addition, due to the saturation of the variation amplitude at relatively low $q\sim 10$, it is not very sensitive even to considerable changes in $\nu$, $\rho_0$
or $m$. 

To sum up, we have shown that microlensing needs to be taken into account in the interpretation of the power spectra and variability amplitudes of various 
sources. These predictions are well substantiated in the case of the Galaxy, and therefore timing properties of pulsars and X-ray binaries has the potential
to reveal properties of the microlensing population of the Galaxy. For cosmological distances, our predictions are weaker because the knowledge of the microlensing
population is less certain and the model of uniformly distributed lenses is unlikely to be applicable in this case. When this difficulty is overcome, one can
attempt to study the cosmological microlensing population via this method.

{\it Acknowledgements.} AVT is supported by IPRS and IPA from the University of Sydney. We thank Rodrigo Gil-Merino for useful comments regarding the
manuscript and the referee for suggestions which lead to a considerable clarification of the presentation. AVT also acknowledges owing drinks 
to J. Berian James for the effort he put into the work on the English of this paper.

\appendix
\section{Unperturbed basis functions and initial conditions}
\label{basisappendix}
A complete eigensystem of solutions to the propagation problem for one-ray statistics was found in Paper I:
\begin{eqnarray}
\lefteqn{ M^{(1)}_{{\bf k}\omega n m}=\zeta'_{n m}(\btau)\frac{\exp\left[-\mathrm{i}\left({\bf k}\cdot{\bf r}+\omega t\right)\right]}{(2\pi)^{3/2}} , } \label{firstordereigenfunctions}
\end{eqnarray}
%\begin{eqnarray}
%\lefteqn{ M^{(1) \dag}_{{\bf k}\omega n m}=\zeta^\dag_{n m}(\tau)\frac{\exp\left[-\mathrm{i}\left({\bf k}\cdot{\bf r}+\omega t+n\chi+\frac{c}{\omega}{\bf k}\cdot\btau \right)\right]}{4\pi^2} , } \label{firstordereigenfunctionscons}
%\end{eqnarray}
\begin{equation}
\lambda_{{\bf k}\omega n m}=-\frac{\mathrm{i}c}{2\omega}{\bf k}^2+\mathrm{i}\frac{\omega}{c}\sqrt{\mathrm{i}\kappa(\omega)}\left(1+|n|+2m\right) , \label{firstordereigenvalues}
\end{equation}
where ($\chi: \btau=\tau {\bf e}_\chi$)
\begin{equation}
\zeta'_{n m}(\btau)=\frac{\exp[-\mathrm{i}(n\chi+\frac{c}{\omega}{\bf k}\cdot\btau)]}{\sqrt{2\pi}}\,\zeta_{nm}(\tau) \label{zetaprimenmbody} 
\end{equation}
and
\begin{eqnarray}
\lefteqn{\zeta_{n m}(\tau)=\sqrt{\frac{2m!\sqrt{\mathrm{i}\kappa(\omega)}}{(m+|n|)!}}\left(\sqrt[4]{\mathrm{i}\kappa(\omega)}\tau\right)^{|n|} }\label{zetanmdef} \\
& & \hspace{10mm}\times \exp\left[-\frac{1}{2}\sqrt{\mathrm{i}\kappa(\omega)}\tau^2\right] L^{|n|}_m\left(\sqrt{\mathrm{i}\kappa(\omega)}\tau^2\right) \nonumber.
\end{eqnarray} 
The `oscillator constant' $\kappa$ in the above expression is defined as
\begin{equation}
\kappa(\omega)\equiv\frac{c\nu}{2\omega}||\mathcal{L}||\langle\bbeta^2\rangle \label{oscillatorconstantdef}
\end{equation}
and, according to our convention, when a square or quartic root of $\mathrm{i}\kappa$ is calculated, one takes the value in the complex plane closest to the positive ray of the real axis.
The conjugate basis $M^{(1)\dag}_{{\bf k} \omega n m}$ can be obtained by complex-conjugating $\zeta_{nm}(\tau)$ function only. See Paper I for details. 
%is the `oscillator constant' defined in Paper I\footnote{Recall a rule from Paper I to select the appropriate value of the square root in~(\ref{firstordereigenvalues}, \ref{zetanmdef}): the one closest to the positive ray of the real axis is chosen.}.
%Because all `non-Hermitecity' of the problem is in the complex-valued potential, $\zeta^\dag$ is just a complex conjugate 
%of $\zeta$, so that when Hermitean conjugation is performed on $M^{(1)}$ only the exponential factor changes phase. 

All possible combinations of the form $M_a({\bf r}_1, {\bf r}_2,t_1, t_2, \btau_1, \btau_2)=M^{(1)}_{a_1}({\bf r}_1, t_1, \btau_1) M^{(1)}_{a_2}({\bf r}_2, t_2, \btau_2)$ 
and associated sums $\lambda_a=\lambda^{(1)}_{a_1}+\lambda^{(1)}_{a2}$ with a composite index $a=({\bf k}_{1,2}, \Omega\mp\omega/2, m_{1,2}, n_{1,2})$
form a dense set of solutions to the unperturbed problem~(\ref{startperturbations}) with zero correlation term.
In this basis, the initial conditions for a synchronous isotropic source of the form~(\ref{M02}) can be written down as:
\begin{eqnarray}
\lefteqn{(M^\dag_{a}, M^0_2)=\frac{16\pi c^2}{\omega^2D_s^2}\frac{\overline{g({\bf k}_1)g({\bf k}_2)} P(\omega)}{|1+\alpha|^2\sqrt{2\kappa(\omega)}}\delta^0_{n_1}\delta^0_{n_2} } \label{c02si} \\
& &\hspace{5mm}\times\exp\left[-\mathrm{i}\frac{c}\omega\left({\bf k}_2^2-{\bf k}_1^2\right)D_s\right]\left(\frac{\alpha-1}{\alpha+1}\right)^{m_1}\left(\frac{\alpha^*-1}{\alpha^*+1}\right)^{m_2} \nonumber
\end{eqnarray}
with
\begin{equation}
\alpha\equiv\frac{c(1-\mathrm{i})}{\omega D_s\sqrt{\kappa(\omega)}} \label{ap}
\end{equation}

\section{Flux functionals}
\label{fluxfunctionalsappendix}
In order to obtain the expectation value of the flux second moment one needs to project the intensity moment vector $M_2$ onto $\btau=0$:
\begin{equation}
F_2=\left.M_2\right|_{\btau_{1,2}=0} \label{fluxpro}
\end{equation}
As explained in the main text, this projection leaves intact the spatial and temporal sectors and therefore for ${\bf k}_{1,2}={\bf K}\mp{\bf k}/2$, $\omega_{1,2}=\Omega\mp\omega/2$, Fourier
modes of the flux are given in $M_a$ basis as a sum
\begin{equation}
F_2\left({\bf K}, {\bf k}, \Omega, \omega\right)=\sum c_a(D)\mathrm{e}^{\lambda_aD} \left.\zeta_a\right|_{\btau_{1,2}=0} \label{f2pro}
\end{equation}
with respect to angular modes only. 

In particular, since $\zeta'_{mn}(\btau=0)=\delta^0_n\sqrt{\sqrt{\mathrm{i}\kappa(\omega)}/\pi}$, 
one has for $\Omega=0$:
\begin{equation}
F_2({\bf K}, {\bf k}, \omega; D)=\frac{\sqrt{2\kappa(\omega)}}{\pi}\sum\limits_{m_{1,2}} c_{{\bf K} {\bf k}\omega m_{1,2}}(D)\mathrm{e}^{\lambda_{{\bf K} {\bf k} \omega m_{1,2}}D} \label{fluxharmonicsOmega0}
\end{equation}
Note, that $n_{1,2}\not=0$ can be ingnored.

These moments can be analogously split into the zeroth-order contribution
\begin{eqnarray}
\lefteqn{F^0_2({\bf K}, {\bf k},\omega; D)=\frac{\sqrt{2\kappa(\omega)}}{\pi}\sum\limits_{m_{1,2}} c^0_{{\bf K} {\bf k} \omega m_{1,2}}\mathrm{e}^{\lambda_{{\bf K} {\bf k} \omega m_{1,2}}D} } \label{F20}
\end{eqnarray}
and a term due to the correlated deflections (see~(\ref{c1a}))
\begin{eqnarray}
\lefteqn{F^1_2({\bf K}, {\bf k},\omega; D)= } \label{F21} \\
& & \frac{\sqrt{2\kappa(\omega)}}{\pi}\int\limits_0^D\mathrm{d}D'\, \sum\limits_{m_{1,2}}\sum\limits_{a'} \mathrm{e}^{\lambda_{b}(D-D')} C_{b a'}c^0_{a'}\mathrm{e}^{\lambda_{a'} D'} ,  \nonumber
\end{eqnarray}
where $b=({\bf K}, {\bf k}, \omega, m_{1,2})$.

%The zeroth-order term describes the overall decay of perturbation due simply to the decay of first moments; it is independent 
%of correlations and may be thought of as a kind of `scaling' of the power spectrum -- it does not introduce any mixing of modes.
%On the contrary, (\ref{F21}) describes a non-trivial `redistribution' and growth of power in the spectrum which arises from 
%the correlated deflections.

The summation above is, in general, rather difficult to perform. However, significant simplification occurs for an isotropic 
and synchronous source of form~(\ref{c02si}). First of all, initial conditions for isotropic sources are proportional to $\delta^0_{n'}$, hence one
does not need to perform summation in $n'$ (these components do develop for $D>0$ but do not contribute to the flux, according to~(\ref{fluxharmonicsOmega0})). 
Second, the set $c^0_b$ in this case is completely defined 
by it spatial profile $\overline{g({\bf K'}-{\bf k'}/2)g({\bf K'}+{\bf k'}/2)}$
and power spectrum $P(\omega')$; hence, due to the linearity of the problem there should be coefficients $F_{dd'}(D)$ such that:
\begin{equation}
F({\bf K}, {\bf k}, \omega; D)=\sum\limits_{d'} F_{dd'}(D) P(\omega')\overline{g({\bf k'}_1)g({\bf k'}_2)} \label{Fdddef}
\end{equation}
with $d=({\bf K}, {\bf k}, \omega)$; here we use the summation instead of the integration sign to avoid unnecessarily long notation.

Noting that for sources described by~(\ref{c02si})
\begin{equation}
\mathrm{e}^{\lambda_{d m_1 m_2}D}=\mathrm{e}^{\lambda_d D}\eta_d(D)^{m_1} (\eta^*_d(D))^{m_2} \label{elambdamm}
\end{equation}
\begin{equation}
c^0_{d m_1 m_2}=P(\omega)\overline{g({\bf k}_1)g({\bf k}_2)}\, c^0_d\, \xi^{m_1} (\xi^*)^{m_2}\label{c02sidmm}
\end{equation}
with
\begin{equation}
\eta_d(D)=\exp\left[-\frac{\omega}{c}\sqrt{\kappa(\omega)}(1+\mathrm{i})D\right] ,\label{etadef}
\end{equation}
\begin{equation}
\xi=\frac{\alpha-1}{\alpha+1} , \label{xidef}
\end{equation}
\begin{equation}
c^0_d=\frac{16\pi c^2}{\omega^2D_s^2}\frac{\exp\left[-\mathrm{i}\frac{2c}\omega({\bf K}\cdot{\bf k})D_s\right]}{|1+\alpha|^2\sqrt{2\kappa(\omega)}}\label{c02sid}
\end{equation}
and
\begin{equation}
\lambda_d=-\mathrm{i}\frac{2c}{\omega}\left({\bf K}\cdot{\bf k}\right)-\frac{\omega}{c}\sqrt{\kappa(\omega)}, \label{lambdaddef}
\end{equation}
it is easy to see that in the zeroth and first order the coefficients $F_{dd'}$ are given as
\begin{eqnarray}
\lefteqn{F^0_{dd'}=\delta^d_{d'} \frac{\sqrt{2\kappa(\omega)}}\pi c^0_d \mathrm{e}^{\lambda_d D} \sum\limits_{m_{1,2}}(\xi\eta_d(D))^{m_1} (\xi^*\eta^*_d(D))^{m_2}} \label{F0dddef}
\end{eqnarray}
and
\begin{eqnarray}
\lefteqn{F^1_{dd'}=\frac{\sqrt{2\kappa(\omega)}}\pi c^0_{d'} \int\limits_0^D\mathrm{d}D'\, \mathrm{e}^{\lambda_d (D-D')}\mathrm{e}^{\lambda_{d'} D'}} \label{F1dddef} \\
& &\hspace{23mm} \times\sum C_{bb'} (\xi\eta')^{m'_1} (\xi^*\eta'^*)^{m'_2}\eta^{m_1}(\eta^*)^{m_2} \nonumber
\end{eqnarray}
where $\eta=\eta_d(D-D')$ and $\eta'=\eta_{d'}(D')$ and the sum is over $m_{1,2}, m'_{1,2}$ running from 0 to infinity.

Summation in~(\ref{F0dddef}) can be done immediately using $\sum z^m=1/(1-z)$ valid $\forall z\in\mathcal{Z}$ : $|z|<1$ and gives
\begin{eqnarray}
\lefteqn{F^0_{dd'}(D)=} \label{F0dd}\\ 
& &\hspace{1mm} \delta^d_{d'}\frac{16c^2}{\omega^2D_s^2}\frac{\exp\left[-\mathrm{i}\frac{2c}\omega\left({\bf K}\cdot{\bf k}\right)\left(D_s+D\right)-\frac{\omega}c\sqrt{\kappa(\omega)}D\right]}{\left|\alpha +1  - (\alpha-1)\exp\left[-\frac{\omega}c\sqrt{\kappa(\omega)}(1+\mathrm{i})D\right]\right|^2}. \nonumber
\end{eqnarray}

In order to evaluate sum in~(\ref{F1dddef}) we recall that $\hat{C}_0$ is a function of ${\bf r}$ and $t$ only and therefore
the matrix elements
\begin{eqnarray}
\lefteqn{C_{bb'}=-\frac\nu{2}||\mathcal{L}||\langle\bbeta^2\rangle \left(M^\dag_{b'}, \btau_1^T \hat{C}_0\btau_2\,M_b\right) }\label{Cbbeval} \\
& & =\int\mathrm{d}^2\btau_1\,\zeta_{-\omega/2,0, m_1}(\tau_1)\zeta_{-\omega'/2, 0, m'_1}(\tau_1)\frac{\exp(\mathrm{i}{\bf Q}_1\cdot\btau_1)}{2\pi}\nonumber \\
& & \hspace{0mm}\times\int\mathrm{d}^2\btau_2\,\zeta_{+\omega/2,0,m_2}(\tau_2)\zeta_{+\omega'/2, 0, m'_2}(\tau_2)\frac{\exp(\mathrm{i}{\bf Q}_2\cdot\btau_2)}{2\pi}\nonumber \\
& & \hspace{5mm}\times \btau_1^T\hat{\tilde{C}_0}({\bf k} - {\bf k}', \omega-\omega')\btau_2 \left(-\frac\nu{2}||\mathcal{L}||\langle\bbeta^2\rangle \delta({\bf K}-{\bf K'})\right)\nonumber
\end{eqnarray}
where 
\begin{equation}
{\bf Q}_{1,2}\equiv\mp 2c\left(\frac{{\bf K}\mp{\bf k}/2}{\omega}-\frac{{\bf K'}\mp{\bf k'}/2}{\omega'}\right)\label{Q12def}
\end{equation}
and $\hat{\tilde{C}_0}(\Delta{\bf k}, \Delta\omega)$ is defined by~(\ref{C0ftdef}).
Functions $\zeta_{\omega, 0, m}$ are given by~(\ref{zetanmdef}):
\begin{equation}
\zeta_{\omega, 0, m}(\tau)=\zeta^0_\omega L_m\left(\sqrt{\mathrm{i}\kappa(\omega)}\tau^2\right) \label{zeta0m}
\end{equation}
with the coefficient $\zeta^0_\omega=\sqrt{2\sqrt{\mathrm{i}\kappa(\omega)}}\exp\left[-\frac{1}{2}\sqrt{\mathrm{i}\kappa(\omega)}\tau^2\right]$ independent of $m$.

Therefore, one can change the order of summation with respect to $m$ and integration with respect to $\tau$ in~(\ref{F1dddef}) and
make use of the Laguerre polynomials generating function:
\begin{equation}
\frac{1}{1-t}\exp\left[-\frac{xt}{1-t}\right]=\sum\limits_m L_m(x) t^m. \label{Laguerregen}
\end{equation}
to obtain for the sum in the second line of~(\ref{F1dddef})
\begin{eqnarray}
\lefteqn{I(D')=-4\nu||\mathcal{L}||\langle\bbeta^2\rangle\sqrt{\kappa(\omega)\kappa(\omega')}\delta\left({\bf K}-{\bf K}'\right)}\label{IDpdef} \\
& & \times\int\mathrm{d}^2\btau_1\mathrm{d}^2\btau_2\,\frac{\exp\left[-\gamma\tau^2_1/2-\gamma^*\btau^2_2/2\right]}{(1-\eta)(1-\xi\eta')(1-\eta^*)(1-\xi^*\eta'^*)} \nonumber \\
& & \hspace{5mm}\times (2\pi)^{-2}\mathrm{e}^{\mathrm{i}\left({\bf Q}_1\cdot\btau_1\right)}\btau_1^T\hat{\tilde{C}_0}\left({\bf k}-{\bf k}',\omega-\omega'\right)\btau_2\mathrm{e}^{\mathrm{i}\left({\bf Q}_2\cdot\btau_2\right)} \nonumber
\end{eqnarray}
where
\begin{equation}
\gamma=(1-\mathrm{i})\left(\sqrt{\kappa(\omega)}\frac{1+\eta}{1-\eta}+\sqrt{\kappa(\omega')}\frac{1+\xi\eta'}{1-\xi\eta'}\right) . \label{bdef}
\end{equation}

If one now denotes $\varphi_{1,2}$ the directions of ${\bf Q}_{1,2}$ $\varphi_{1,2}:{\bf Q}_{1,2}=Q_{1,2}{\bf e}_{\varphi_{1,2}}$,
angular integration in~(\ref{IDpdef}) yields a product of two first-order Bessel functions and radial integration in $\tau$
can be performed using identity $\int_0^\infty\mathrm{d}\tau\,\tau^2 J_1(Q\tau)\exp(-\gamma\tau^2/2)=Q/\gamma^2\exp(-Q^2/2\gamma)$ \citep{GR}. Thus we arrive at
\begin{eqnarray}
\lefteqn{ I(D')=4\nu\sqrt{\kappa(\omega)\kappa(\omega')}\delta\left({\bf K}-{\bf K}'\right)\tilde{C}\left(\Delta{\bf k},\Delta\omega\right)}\label{IDp} \\
& & \hspace{-5mm} \times \frac{Q_1 Q_2\exp\left[-Q_1^2/2\gamma-Q_2^2/2\gamma^*\right]\cos({\varphi_1}-\alpha)\cos(\alpha-{\varphi_2})}{\gamma^2(1-\eta)(1-\xi\eta')\,\gamma^{*2}(1-\eta^*)(1-\xi^*\eta'^*)}  \nonumber
\end{eqnarray}
and
\begin{eqnarray}
\lefteqn{F^1_{dd'}(D)=\sqrt{\frac{\kappa(\omega)}{\kappa(\omega')}}\frac{16c^2}{\omega'^2D_s^2}\frac{\exp\left[-\mathrm{i}\frac{2c}{\omega'}({\bf K}'\cdot{\bf k}')D_s\right]}{|1+\alpha|^2}} \label{F1dd} \\
& & \hspace{10mm}\times\int\limits_0^D\mathrm{d}D'\, I(D')\exp\left[\lambda_d(D-D')+\lambda_{d'}D'\right] \nonumber
\end{eqnarray}

% and will be the primary subject of our calculations. The sum in this formula is, in general,
%quite difficult to perform. However, for the initial conditions we consider, the coefficients $c^0_a$ decrease rapidly
%with increasing $m$ and, consequently, $\lambda$ (see Paper I). Therefore, we can consider $D$ small enough such that 
%$\exp(\lambda D)\approx 1+\lambda D + \overline{o}(D)$ and therefore
%\begin{equation}
%C_{aa'}(D)\approx D C_{aa'}=-\frac\nu 2 D||\mathcal{L}||\langle\bbeta^2\rangle \left(M^\dag_a,\btau_1^T\hat{C}_0\btau_2 M_{a'}\right) \label{CD1} 
%\end{equation}
%Then, from the linearity of the scalar product and the definition of $c^0_{a'}$, one has
%\begin{equation}
%F^1_2\approx-\frac{\nu\sqrt{2\kappa(\omega)}}{2\pi}D||\mathcal{L}||\langle\bbeta^2\rangle\sum\limits_{m_{1,2}}\left(M^\dag_b,\btau_1^T\hat{C}_0\btau_2 M^0_2\right) \label{F21D}
%\end{equation}

\section{Modification of the temporal power spectrum}
\label{psiappendix}
In this appendix we calculate quantities~(\ref{F1rribody}) and the line profile function $\psi$ in the simple case of a point-like source at the origin of the source plane $g({\bf k})=1$ as seen by a single observer at the origin of its plane ${\bf r}_{1,2}=0$.

Using~(\ref{F0dd}) one obtains, in this situation (care must be taken when calculating the limit of $\kappa\to 0$ because $\alpha$ tends to infinity in this limit):
\begin{eqnarray}
\lefteqn{\mathcal{F}^0(\omega', D)=} \label{F0ddkappao}\\
& & \hspace{-7mm} \frac{4P(\omega')}{D_s^2\left(D_s+D\right)^2}\frac{\exp\left[-\frac\omega c\sqrt{\kappa(\omega)}D\right]}{\left|\alpha +1  - (\alpha-1)\exp\left[-\frac{\omega}c\sqrt{\kappa(\omega)}(1+\mathrm{i})D\right]\right|^2} \nonumber \\
& & \hspace{-7mm} =\frac{P(\omega')}{(D_s+D)^4}+\mathcal{O}\left(\kappa^2\right) . \nonumber
\end{eqnarray}
Therefore, in the leading order, the line profile function $\psi$ is given by the expression~(\ref{F1rrbody}).

The last line of~(\ref{F0ddkappao}) also shows that the zeroth-order correction to the free propagation of flux is of the second
order in $\kappa$ (and therefore $\nu$); as shown below, correction in the first order of perturbation due to correlated deflections 
is~$\mathcal{O}\left(\sqrt{\kappa}\right)$ and therefore represents the dominant term.

To perform integration~(\ref{F1rrbody}) for the first-order correction, we first untilize the $\delta$-function in ${\bf K}-{\bf K}'$ from~(\ref{F1dd})
and then, taking into account that $\tilde{C}$ depends on $\Delta{\bf k}={\bf k}-{\bf k}'$ only, perform a change of integration
variables $\left({\bf K}, {\bf k}, {\bf k}'\right)\rightarrow\left({\bf Q}_1, {\bf Q}_2, \Delta{\bf k}\right)$ according to
\begin{eqnarray}
& &\hspace{-7mm} \lefteqn{{\bf K}=\frac{\omega\omega'}{4c\Delta\omega}\left({\bf Q}_1-{\bf Q}_2\right)} \nonumber\\
& &\hspace{-7mm} \lefteqn{{\bf k}=-\frac{\omega\omega'}{2c\Delta\omega}\left({\bf Q}_1+{\bf Q}_2\right)+\frac{\omega}{\Delta\omega}\Delta{\bf k}} \label{Kkk2QQdk} \\
& &\hspace{-7mm} \lefteqn{{\bf k}'=-\frac{\omega\omega'}{2c\Delta\omega}\left({\bf Q}_1+{\bf Q}_2\right)+\frac{\omega'}{\Delta\omega}\Delta{\bf k}} . \nonumber
\end{eqnarray}
The Jacobian determinant of this transformation
\begin{equation}
\frac{D\left({\bf K}, {\bf k}, {\bf k}'\right)}{D\left({\bf Q}_1, {\bf Q}_2, \Delta{\bf k}\right)}=\left(\frac{\omega\omega'}{2c\Delta\omega}\right)^4 \label{Kkk2QQdkJacobian}
\end{equation}
is independent of integration variables and thus appears as a factor in front of the integrals, and since this is a linear
transformation ${\bf Q}_{1,2}, \Delta{\bf k}$ independently run through entire $\mathcal{R}^2$ planes.

Then, we may rewrite the first-order correction~(\ref{F1rrbody}) as 
\begin{eqnarray}
\lefteqn{\psi^1(\omega, \omega', D)=} \label{P1P} \\
& &\hspace{0mm}\sqrt{\frac{\kappa(\omega)}{\kappa(\omega')}}\left(\frac{\omega\omega'}{\Delta\omega}\right)^4 \frac{(D_s+D)^4}{c^2\omega'^2D_s^2|1+\alpha|^2}\exp\left[-\frac\omega{c}D\sqrt{\kappa(\omega)}\right] \nonumber \\
& &\hspace{-7mm} \times\int\limits_0^D\mathrm{d}D'\,\frac{4\nu\sqrt{\kappa(\omega)\kappa(\omega')}\exp\left[-D'(\frac{\omega'}c\sqrt{\kappa(\omega')}-\frac\omega c\sqrt{\kappa(\omega)})\right]}{(1-\eta)(1-\xi\eta')(1-\eta^*)(1-\xi^*\eta'^*)} \nonumber \\
& &\hspace{-7mm} \times\int\mathrm{d}^2\Delta{\bf k}\,\tilde{C}(\Delta{\bf k},\Delta\omega)\int\mathrm{d}Q_1\,\mathrm{d}Q_2\,\frac{Q_1^2Q_2^2}{\gamma^2\gamma^{*2}}\exp\left[-\frac{Q_1^2}{2\gamma}-\frac{Q_2^2}{2\gamma^*}\right] \nonumber \\
& &\hspace{-7mm} \times\int\mathrm{d}\varphi_1\,\mathrm{d}\varphi_2\,\frac{\cos(\varphi_1-\alpha)\cos(\varphi_2-\alpha)}{4\pi^2} \nonumber \\
& &\hspace{-7mm} \times\exp\left[-\mathrm{i}\frac{2c}{\omega}\left({\bf K}\cdot{\bf k}\right)\left(D-D'\right)-\mathrm{i}\frac{2c}{\omega'}\left({\bf K}'\cdot{\bf k}'\right)\left(D_s+D'\right)\right]; \nonumber
\end{eqnarray}
here we deliberately postpone some obvious cancellations to keep the asymptotic behaviour clear and provide for lens population
parameters dependence on $D$.

Using~(\ref{Kkk2QQdk}) we rewrite the argument of the exponent in the last line in terms of ${\bf Q}_{1,2}$ and $\Delta{\bf k}$,  angular integration yields a product of two Bessel functions and the identity used in deriving~(\ref{IDp}) brings the integral
with respect to ${\bf Q}_{1,2}$ to the following form:
\begin{eqnarray}
\lefteqn{u(\Delta{k})=\left(\frac{2\Delta\omega^2}{\omega\omega'}\right)^2\frac{p^2\Delta k^2}{\left(D_s+D\right)^2}\exp\left[-p\,\mathrm{Re}\gamma\,\Delta k^2\right]} \label{fsk}
\end{eqnarray}
with ($|\gamma|^2=\mathrm{Im}^2\gamma+\mathrm{Re}^2\gamma$) 
\begin{equation}
p\equiv \left(D_s+D\right)^2/\left[\left(\frac{2\Delta\omega^2}{\omega\omega'}+\delta\mathrm{Im}\gamma\right)^2+\delta^2\mathrm{Re}^2\gamma\right] , \label{pdef}
\end{equation}
where
\begin{equation}
\delta\equiv\frac{\omega'(D-D')+\omega(D_s+D')}c .\label{deltadef}
\end{equation}

Function $u(\Delta k)$ is independent of its argument vector direction, and integration of $\tilde{C}(\Delta{\bf k}, \Delta\omega)$
with respect to the angular coordinate produces Infeld (modified Bessel) function $2\pi I_0(v\Delta\omega/\sigma^2\Delta k)$.
Therefore, one has for the integral with respect to $\Delta{\bf k}$:
\begin{eqnarray}
\lefteqn{\int\mathrm{d}^2\Delta{\bf k}\,\tilde{C}(\Delta{\bf k}) u(\Delta k) =\left(\frac{2\Delta\omega^2}{\omega\omega'}\right)^2\frac{2\sqrt{2\pi} m^2 p^2\mathrm{e}^{-\frac{v^2}{2\sigma^2}}}{\sigma\left(D_s+D\right)^2}} \label{dkint} \\
& &\times\int\mathrm{d}\Delta k\,\exp\left[-\frac{1}{2}\left(\frac{\Delta\omega^2}{\sigma^2\Delta k^2}+p\,(\gamma+\gamma^*)\Delta k^2\right)\right] \nonumber \\
& &\hspace{14mm}\times I_0\left(\frac{v\Delta\omega}{\sigma^2\Delta k}\right)\left[J_0\left(\frac{\rho_0}2\Delta k\right) -J_0\left(\Lambda\frac{\rho_0}2\Delta k\right) \right]^2  \nonumber
\end{eqnarray}
The exponential factor decreases rapidly at $\Delta k$ tending to both zero and infinity thus providing necessary convergence.
%One can introduce the value $\Delta k_0$ where the argument of the exponent $-q/2((\Delta k/\Delta k_0)^2+(\Delta k/\Delta k_0)^{-2})$ reaches its maximum value $-q$:
%\begin{equation}
%\Delta k_{max}^2=\frac{\Delta\omega}{\sigma}\frac{1}{\sqrt{p(\gamma+\gamma^*)}}, \label{dkmaxdef}
%\end{equation}
%\begin{equation}
%q=\frac{\Delta\omega}{\sigma}\sqrt{p\,(\gamma+\gamma^*)} \label{qdef}
%\end{equation}
%and neglect integration outside the range between $\mathcal{O}(\Delta k_{max}\sqrt{q})$ and $\mathcal{O}(\Delta k_{max}/\sqrt{q})$.

It is reasonable now to introduce some approximations. It was shown in Paper I that in physically reasonable
situations $\sqrt{\kappa(\omega)}\ll 1$. We can use this fact and expand $\eta$, $\xi$, $\eta'$ etc. in its power series retaining
only the few leading terms as necessary.

First of all, we have for the denominator in $I(D')$
\begin{equation}
|1-\eta|^2=2\kappa(\omega)\left(\frac{\omega}c\left(D-D'\right)\right)^2+\overline{o}\left(\kappa\right) \label{a1meta2s}
\end{equation}
and
\begin{equation}
|1-\xi\eta'|^2=2\kappa(\omega')\left(\frac{\omega}c\left(D_s+D'\right)\right)^2+\overline{o}\left(\kappa\right) \label{a1mxietap2s}
\end{equation}
and similarly
\begin{equation}
\frac{1}{|1+\alpha|^2}=\frac{\omega'^2D_s^2\kappa(\omega')}{2c^2} .\label{alphap1as}
\end{equation}
For $\gamma$ the first term in expansion is purely geometrical and imaginary while the first real term is proportional to $\kappa$:
\begin{eqnarray}
\lefteqn{\gamma=-2\mathrm{i}c\frac{\omega(D-D')+\omega'(D_s+D')}{\omega\omega'(D-D')(D_s+D')}+\mathrm{i}\mathcal{O}(\kappa^2)}  \label{gammas}\\
& &  +\omega\kappa(\omega)\frac{D_s+D}{3c}\left[1-\frac{D_s^3}{(D_s+D)(D_s+D')^2}\right] + \mathcal{O}(\kappa^3),  \nonumber
\end{eqnarray}
where we have used $\omega\kappa(\omega)=\omega'\kappa(\omega')$ from~(\ref{oscillatorconstantdef}).

Expanding $\Delta\omega^2$ and using the above formulae one finds that at this level of approximation $p$ is purely
geometrical in the leading order:
\begin{equation}
p=\frac{(D-D')^2(D_s+D')^2}{(D_s+D)^2+\mathcal{O}(\kappa^2)}=(D_s+D)^2t^2(1-t)^2+\mathcal{O}(\kappa^2) \label{ps}
\end{equation}
with dimensionless distance coordinate $t=(D_s+D')/(D_s+D)$ introduced; as $D'$ changes from $D_s$ to $D$, the corresponding
range for $t$ is from $s=D_s/(D_s+D)$ to $1$. 

Now, assuming that parameters $\nu, m, \sigma, v$ are independent of $D'$, we use~(\ref{oscillatorconstantdef}) and 
$\langle\bbeta^2\rangle=2\pi||\mathcal{L}||^{-1}m^2\ln\Lambda$, introduce the frequency difference scale
\begin{equation}
\Delta\omega_0=\frac{2\sigma}{\rho_0} , \label{domega0def}
\end{equation}
dimensionless variables
\begin{equation}
x=\frac{\rho_0}2\Delta k, \label{xdef}
\end{equation}
\begin{equation}
y=\frac{\Delta\omega}{\Delta\omega_0} \label{ydef}
\end{equation}
and an amplitude parameter
\begin{eqnarray}
\lefteqn{q=\sqrt{\frac{2\omega\kappa(\omega)}{3c}}\frac{2\left(D_s+D\right)^{3/2}}{\rho_0}} \label{qdef}\\
& &\hspace{-5mm} =\frac{m}{\rho_0}\sqrt{\frac{8\pi\ln\Lambda\,\nu\left(D_s+D\right)^3}{3}} , \nonumber
\end{eqnarray}
we write down the following approximation for the ratio~(\ref{P1P}):
\begin{eqnarray}
\lefteqn{\psi(\Delta\omega, D)=\frac{12\exp\left[-v^2/2\sigma^2\right]}{\sqrt{2\pi}\ln\Lambda\,\Delta\omega_0} }\label{P1Ps} \\
& & \hspace{-3mm} \times q^2\int\limits_s^1\mathrm{d}t\,\int\limits_0^\infty\mathrm{d}x\,t^2(1-t)^2\exp\left[-\frac{x^2q^2}{2}(1-t)^2(t^2-s^3)\right] \nonumber\\
& & \hspace{0mm} \times\exp\left[-\frac{y^2}{2x^2}\right] I_0\left(\frac v\sigma \frac y x\right)\left[J_0\left(x\right)-J_0\left(\Lambda x\right)\right]^2+\overline{o}(\sqrt\nu). \nonumber
\end{eqnarray}
-- as expected, it is~$\mathcal{O}(\sqrt{\nu})$.

\begin{figure}
\hspace{0cm}
\includegraphics[width=80mm, angle=0]{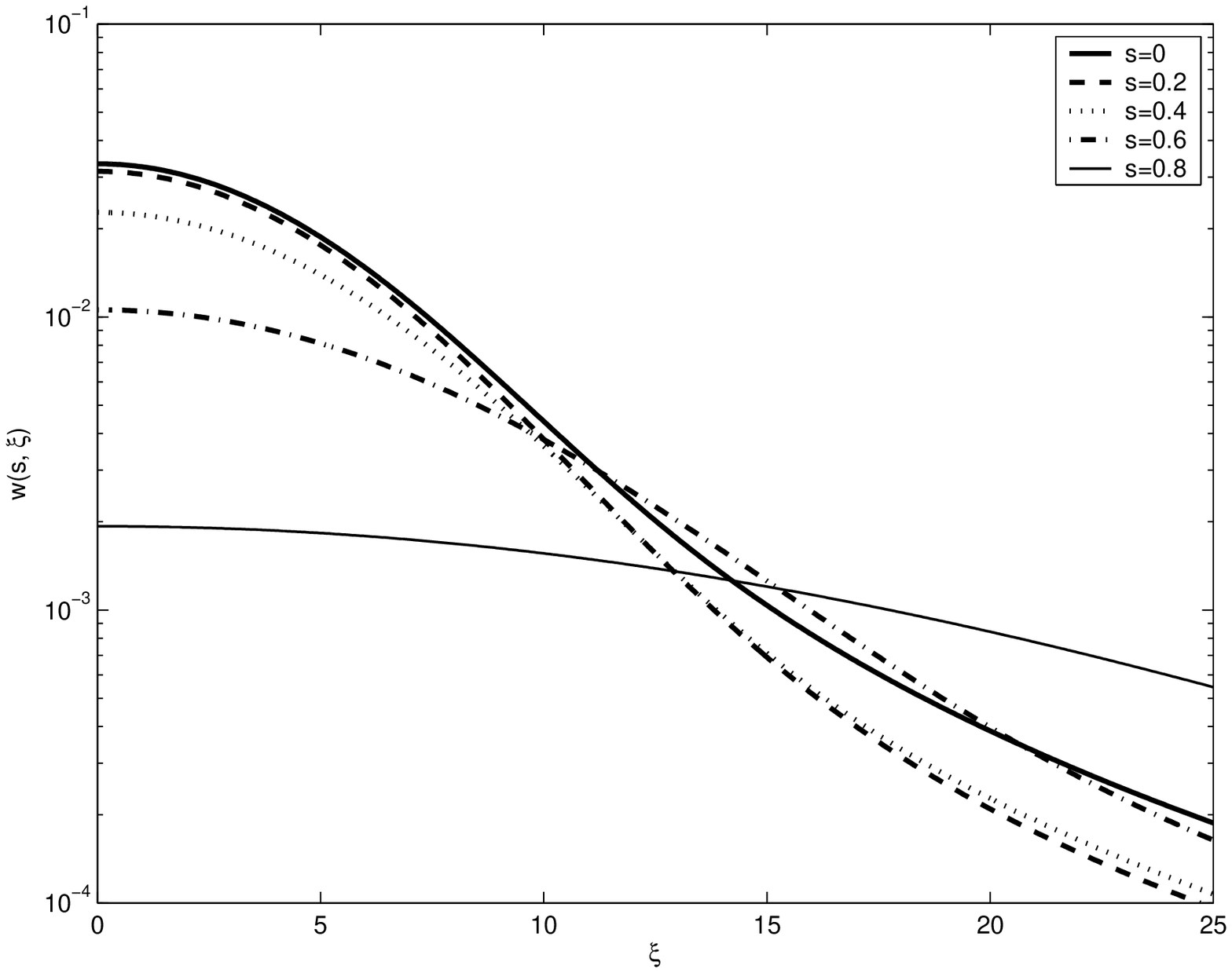}
\caption{Auxiliary function $w(s, \xi)$ defined by the integral~(\ref{wdef}). Behaviour of $w$ as a function of $\xi$ is displayed
for five different values of $s$ -- $0$, $0.2$, $0.4$, $0.6$ and $0.8$; $w(s, \xi)\propto\xi^{-3}$ for large $\xi$.}
\label{wfig}
\end{figure}

From this point, integration needs to be numerical, and is, perhaps, most easily done by first calculating the 
integral of the function on the second line of~(\ref{P1Ps}) with respect to $t$ and then performing calculation in $x$.
This approach yields the expression~(\ref{P1PPhibody})
with the scaled line profile function
\begin{eqnarray}
\lefteqn{\Psi\left(y, s, \frac v\sigma, q\right)\equiv} \label{Psidef} \\
& & \hspace{-5mm} q^2\int\limits_0^\infty\mathrm{d}x\, w\left(s, xq\right)\exp\left[-\frac{y^2}{2x^2}\right] I_0\left(\frac v\sigma \frac y x\right)\left[J_0\left(x\right)-J_0\left(\Lambda x\right)\right]^2  \nonumber 
\end{eqnarray}
defined through an auxiliary function
\begin{equation}
w(s, \xi)\equiv\int\limits_s^1\mathrm{d}t\,t^2(1-t)^2\exp\left[-\frac{1}2\xi^2(1-t)^2(t^2-s^3)\right] \label{wdef}
\end{equation}
The behaviour of $w(s, \xi)$ as a function of $\xi$ is shown in Figure~\ref{wfig} for a few values of $s$; $w(s, \xi)\propto \xi^{-3}$
for $\xi\gg 1$. Since $\Lambda\gg 1$ we can safely neglect the second Bessel function in the square brackets when evaluating~(\ref{Psidef}).

\label{lastpage}
\end{document}